\documentclass[sigconf]{acmart} 
\usepackage[english]{babel}
\usepackage{blindtext}
\usepackage{lineno}
\usepackage{cuted}
\usepackage{cmbright}
\usepackage{enumitem}
\usepackage{fancyhdr}

\DeclareUnicodeCharacter{2606}{ }

\newcommand{\avela}[1]{AVELA}
\newcommand{\uw}[1]{UW}
\newcommand{\wa}[1]{Washington}
\newcommand{\sea}[1]{Seattle metropolitan area}

\renewcommand\footnotetextcopyrightpermission[1]{} 
\setcopyright{none}
\acmConference[]{}
\acmYear{}
\copyrightyear{}

\begin{document}


\title[\avela{}]{AVELA - A Vision for Engineering Literacy \& Access: Understanding Why Technology Alone Is Not Enough}

\author{Kyle Johnson}
\authornote{Equal and primary contribution first authors.}
\authornote{Also with the University of Washington}
\email{chair@avelaccess.org}
\orcid{0000-0002-0443-9892}
\affiliation{
  \institution{AVELA - A Vision for Engineering Literacy \& Access}
  \streetaddress{P.O. Box ????}
  \city{Seattle}
  \state{Washington}
  \country{USA}
  \postcode{98195}
}

\author{Vicente Arroyos}
\authornotemark[1]
\authornotemark[2]
\email{coordinator_uw@avelaccess.org}
\orcid{XXXX-XXXX-XXXX}
\affiliation{
  \institution{AVELA - A Vision for Engineering Literacy \& Access}
  \streetaddress{P.O. Box ????}
  \city{Seattle}
  \state{Washington}
  \country{USA}
  \postcode{98195}
}
\author{Celeste Garcia}
\authornotemark[1]
\authornotemark[2]
\email{celesg@uw.edu}
\orcid{XXXX-XXXX-XXXX}
\affiliation{
  \institution{AVELA - A Vision for Engineering Literacy \& Access}
  \city{Seattle}
  \state{Washington}
  \country{USA}
  \postcode{98195}
}
\author{Liban Hussein}
\authornotemark[1]
\authornotemark[2]
\email{libanh@uw.edu}
\orcid{0000-0002-2298-5424}
\affiliation{
  \institution{AVELA - A Vision for Engineering Literacy \& Access}
  \city{Seattle}
  \state{Washington}
  \country{USA}
  \postcode{98195}
}
\author{Aisha Cora}
\authornotemark[1]
\authornotemark[2]
\email{aishac16@uw.edu}
\orcid{0009-0007-6245-0989}
\affiliation{
  \institution{AVELA - A Vision for Engineering Literacy \& Access}
  \city{Seattle}
  \state{Washington}
  \country{USA}
  \postcode{98195}
}

\author{Tsewone Melaku}
\email{tsewonemelaku@gmail.com}
\orcid{XXXX-XXXX-XXXX}
\affiliation{
  \institution{AVELA - A Vision for Engineering Literacy \& Access}
  \city{Seattle}
  \state{Washington}
  \country{USA}
  \postcode{98195}
}
\author{Jay L. Cunningham}
\email{jaylcham@uw.edu}
\orcid{0003-2446-8022}
\affiliation{
  \institution{University of Washington}
  \city{Seattle}
  \state{Washington}
  \country{USA}
  \postcode{98195}
}

\author{R. Benjamin Shapiro}
\email{rbs@cs.washington.edu}
\orcid{XXXX-XXXX-XXXX}
\affiliation{
  \institution{University of Washington}
  \city{Seattle}
  \state{Washington}
  \country{USA}
  \postcode{98195}
}

\author{Vikram Iyer}
\email{vsiyer@uw.edu}
\orcid{XXXX-XXXX-XXXX}
\affiliation{
  \institution{University of Washington}
  \city{Seattle}
  \state{Washington}
  \country{USA}
  \postcode{98195}
}
\renewcommand{\shortauthors}{Kyle and Vicente, et al.}

\begin{abstract}
Unequal technology access for Black and Latine communities has been a persistent economic, social justice, and human rights issue despite increased technology accessibility due to advancements in consumer electronics like phones, tablets, and computers. We contextualize sociotechnical access inequalities for Black and Latine urban communities and find that many students are hesitant to engage with available technologies due to a lack of enticing support systems. We develop a holistic student-led STEM engagement model through \avela{} leveraging near-peer mentorship, experiential learning, mentor embodied community representation, and culturally responsive lessons. We conduct 24 semi-structured interviews with college \avela{} members, analyze 171 survey responses from \avela{}'s secondary school class participants, and apply autoethnographic analysis. We evaluate the model's impact after 4 years of mentoring 200+ university student instructors in teaching to 2,500+ secondary school students in 110+ classrooms. We identify access barriers and provide principled recommendations for designing future STEM education programs.

\end{abstract}




\begin{teaserfigure}
  \centering
  \includegraphics[width=0.95\linewidth]{Figures/TeaserFigure.pdf}
  \caption{Visualization of the student-led STEM engagement model showcasing the \avela{}'s cyclical impact on secondary school students, college undergraduate and graduate students, as well as research projects and tools. }
  \Description{Four images that showcase the cyclical nature of the \avela{} model. First image of secondary school students pointing to becoming undergraduate school students, which then points to becoming graduate school students, which then points to creating research projects & tools based off their graduate school education, which all goes back to teaching secondary school students for the cycle to be repeated.}
  \label{fig:teaser}
\end{teaserfigure}


\maketitle

\thispagestyle{fancy}
\fancyhead{}
\fancyhead[L]{}
\fancyhead[C]{This is the author's version of the work. It is posted here for personal use, not for redistribution.}
\fancyhead[R]{}
\setlength{\headheight}{25pt}
\pagestyle{plain} 

\newpage
\section{Introduction}

During the COVID-19 pandemic, schools around the world were reminded of the persistence of the digital divide as the quality of a student's education became more clearly tied to the access and usability of educational technologies~\cite{digital-divide,covid-education1,covid-education2}. Technology access typically refers to the hardware and software services physically available for use by an individual~\cite{nsf-access}. This digital divide is further exacerbated within science, technology, engineering, and mathematics (STEM) fields, as the technology requirement for learning about these topics is usually higher~\cite{levg-tech-stem}. 

Prior work has brought attention to and analyzed many essential factors in improving STEM education for Black and Latine students in urban communities. Approaches like assets-based community cultural wealth~\cite{ccw,ccw-chi}, transformative justice~\cite{transformative-justice}, experiential and service learning~\cite{sl-chi,hotchkins2019first, patterson2021you, co-curricular}, culturally responsive teaching~\cite{cr-chi,culturally-responsive}, as well as student-driven and research-based initiatives~\cite{alias2016student,nsbe-27280,LUEDERITZ2016229} have proven to benefit the quality of education for Black and Latine students. In the U.S. specifically, Black and Latine communities have faced a long history of discrimination and injustice, from slavery and Jim Crow segregation to ongoing issues around economic inequality, educational funding gaps, racial profiling, voter suppression, and disparities in healthcare treatment. This racial climate in the U.S. leaves Black and Latine students facing unique challenges when trying to work in STEM fields, challenges that are too often insurmountable for aspiring scholars. 

Years of research show consistent underrepresentation of Black and Latine individuals in STEM careers~\cite{eeoc2022, bw-stem, kept_out, digital_racial_gap}. Why does this gap persist despite strong interest from Black and Latine communities to participate in STEM fields~\cite{nsbe-27280,nsbe-demographics,shpe-demographics}? Beyond having access to physical technologies and a broad interest in these topics, many students remain unaware of how to effectively interact with these technologies to either benefit their communities or get access to new career pathways. Moreover, many students feel unwelcome in academic spaces which can lower participation~\cite{patterson2021you}, and those who persist often experience feelings of imposter syndrome~\cite{imposter-syndrome,is-hs}. One can physically have access to technologies but still be discouraged from learning about how to use them if the learning environment is perceived as unapproachable. 
Thus, we must consider access from a sociotechnical perspective.

Numerous shared experiences with this disparity in sociotechnical access to STEM fields inspired students at \uw{}
to create a student-led organization for increasing Black and Latine students' interest and persistence in STEM careers, now known as \avela{}.
From the start, these students' experiences showed that physical access to technology alone was not enough to address persistent disparities, which compelled them to employ an assets-based community cultural wealth approach to engaging with other Black and Latine youth~\cite{ccw}. In this paper, we refer to the combination of specific methodologies implemented by the \avela{} as a student-led STEM engagement model, shown in Fig~\ref{fig:teaser}.

We present \avela{}'s holistic model to demonstrate that while computing and technology solutions alone are not enough, combining these with a human centered sociotechnical approach can successfully reduce access inequalities for Black and Latine urban communities in the U.S. Since the summer of 2019, \avela{} has used their mentorship model to engage 200+ university student instructors in teaching culturally responsive STEM classes in 112 classrooms and community centers, reaching approximately 2,500 secondary school students and bringing together more than 50 community partners and collaborators. To evaluate the impact of the \avela{}'s model, we conduct 24 semi-structured interviews with college \avela{} members, analyze 171 pre- and post-class survey responses from the \avela{}’s secondary school class participants, and present our own reflection as the founders of \avela{}. 

This study makes the following specific contributions to understanding the barriers Black and Latine students face when engaging with computing and technology as well as providing concrete solutions for designing effective outreach models to address them: 
\begin{itemize}[topsep=0pt]
\item
\textbf{Contexualizing sociotechnical access inequalities.} 
We contexualize the digital divide for Black and Latine urban students through a qualititive and autoethnographic analysis, looking beyond physical access to computing and technology. We found that while most interview participants had access to computers and the internet, they only used them for basic tasks like word processing and were hesitant to explore the technologies further due to them being unapproachable and not knowing how else to interact with them.

\item\textbf{Propose a student-led STEM engagement model.} We outline three key program design principles based on \avela{}'s student-led STEM engagement model: A) Near-peer mentorship and interactive \& experiential learning. B) Mentor embodied community representation and culturally responsive lesson plans. C) Student-led community engagement leveraging academic research and mentors. We observe that a holistic program model combining all of these factors is critical for supporting a broad range of student backgrounds. 

\item\textbf{Evaluate the impact of our proposed design through a multimethod study.} 
We found that representative and culturally responsive mentorship can help increase the approachability of STEM spaces, and that skill development and career enhancement through service learning can entice students to participating in STEM programs. 

\item\textbf{Provide programmatic recommendations for academic stakeholders.} We provide insight on the unique aspects of \avela{}'s model, offer an explanation for why it has scaled successfully, and provide recommendations for academic stakeholders, like administrators at other universities, to consider when designing STEM education programs for Black and Latine urban communities.
\end{itemize} 

\noindent This study confirms our design intuitions that \avela{}'s programmatic methodologies are having a meaningful impact on mitigating the digital divide for our target population. Through our qualitative and autoethnographic analysis we surfaced a more nuanced understanding on the digital divide. In conjunction with these analyses, our quantitative data analysis rigorously evaluated the impact of \avela{}'s student-led STEM engagement model. Rooted in the qualitative, quantitative, and autoethnographic findings we then provide recommendations to academic stakeholders for program design principles that should be focused on when designing a STEM outreach program for Black and Latine urban communities.
\section{Background and related work}
\subsection{Digital Divide- A National and Local Context}
Advancements in technology for computing and clean energy to healthcare and manufacturing will be key in addressing the global challenges of the next few decades, necessitating a qualified STEM workforce. Projections show that by 2060, Black and Latine youth will comprise nearly half of all U.S. school-age children. However, these students are currently severely underrepresented in STEM and will only continue to fall further behind without support~\cite{kept_out, digital_racial_gap}. 
Prior works have explored a myriad of frameworks and models for addressing inequities in sociotechnical access. These approaches include social design experiments to validate educational interventions in practice and promote social equity~\cite{equity-design}, integrational co-design~\cite{co-design-action,co-design-process}, leveraging family-school-community leadership to be more culturally responsive~\cite{shanee1}, service learning~\cite{service-learning}, leveraging cultural context for learning~\cite{chicana-math-methodology}, side by side learning~\cite{disruptive-teaching}, in-school and out-of-school programs with role models ~\cite{Chen2023} and generative computing for enhancing connections between Black heritage and CS~\cite{aa-cosmetology}. These approaches provide promising frameworks for addressing STEM access inequalities, however there remains a need for a many-method solution grounded in the lived experiences and articulated needs of the target communities to be both implemented and evaluated. In this work we develop a many-methods solution combining numerous proven pedagogical practices to provide educational STEM interventions for Black and Latine youth and show that this combined methods approach is uniquely successful in maximizing student engagement, learning, and impact.

\subsection{Barriers Beyond Physical Access}
The gap in STEM program enrollment for URM students is not due to a lack of interest from the students or their families~\cite{nsbe-27280,nsbe-demographics,shpe-demographics,margolis2017stuck}. Rather, Black and Latine students face significant barriers to enrolling in and graduating from college due to decades of systemic inequality. While research and policy initiatives on this topic have often framed the digital divide as a technology access issue, a more nuanced perspective is necessary~\cite{disalvo2017participatory}. DiSalvo describes how individual and family monetary needs can outweigh students' interest in pursuing STEM education, and illustrates how educational programs that provide for both financial need and subject matter interest can create new inroads to STEM for marginalized students\cite{disalvo2014saving}. For example, \uw{} enrollment data indicates that over the past 5 years, an average of 48\% of self-identifying URM students are Pell-elligible~\cite{pell_urm_all_UW}, a federally recognized indication of financial need~\cite{pell}. Moreover, barriers to STEM participation have compounded over time to form a mentoring gap; when students do not see people who look like them in STEM, it is difficult for them  see STEM as a career pathway for themselves~\cite{role-models}. Educational interventions that emphasize near-peer mentoring have shown promise to address this challenge, but bootstrapping such programs remains a challenge.

In order to repair the harm created by institutions, policies, and practices that have systematically excluded Black and Latine youth from computer science, an intersectional, transformative justice approach must be taken~\cite{transformative-justice}. A transformative justice approach creates alternatives structures that center the relationships and communities of those who have been historically excluded from educational institutions and targets of state violence. Transformative justice targets specific laws, policies, and practices by creating counter-structures and co-designed community-based alternatives. The field of human-computer interaction has begun to incorporate transformative justice methods into the design values and methods of new technologies~\cite{tj-chi,aa-cosmetology}. These prior works highlight the importance of proper framing when trying to mitigate disparities and foster design justice~\cite{costanza2018design} in technology accessibility. Informed by this, \avela{} takes an assets-based community cultural wealth approach which centers student-led initiatives similar to organizations like the National Society of Black Engineers (NSBE) and Society of Hispanic Professional Engineers (SHPE). 


\begin{figure*}[t]
  \centering\includegraphics[width=1.0\linewidth]{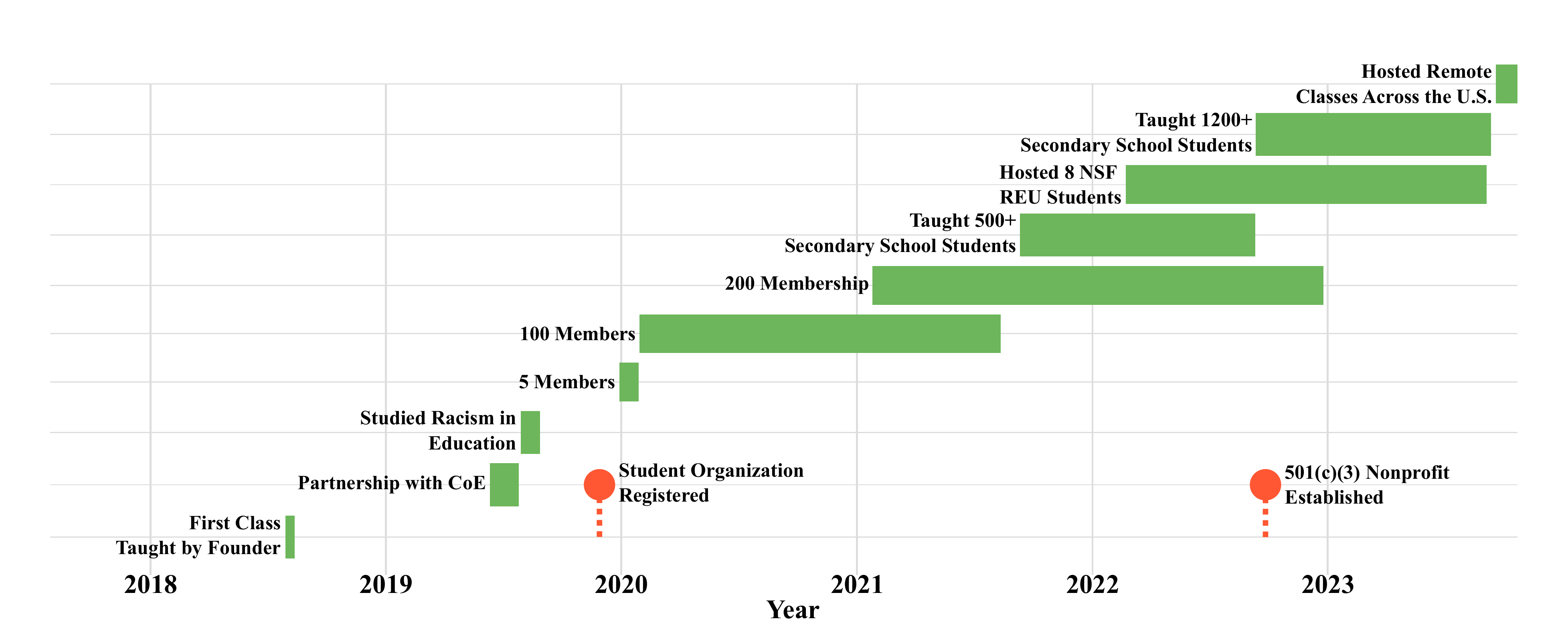}
  \caption{\avela{} timeline with key milestones highlighted by year. \textbf{A)} First \avela{} class taught by founder. \textbf{B)} \avela{} founders partner with \uw{} engineering department to teach 3 summer classes. \textbf{C)} Founder meets secondary student named \avela{} while studying racism in education systems in Cape Town, South Africa. \textbf{D)} \avela{} Submits \uw{} registered student organization documents. \textbf{E)} Membership grows from 2 to 5 students. \textbf{F)} Membership grows from 5 to 100 students. \textbf{G)} Membership grows from 100 to 200 students. \textbf{H)} Members teach 500+ secondary school students. \textbf{I)} \avela{} helps host 8 NSF REU students. \textbf{J)} \avela{} establishes 501(c)(3) nonprofit. \textbf{K)} Members teach 1200+ secondary school students. \textbf{L)} \avela{} co-hosts remote classes in cities across the U.S.}
  \Description{Figure showcasing timeline of growth for \avela{} from 2018 to 2023.}
  \label{fig:avela-timeline}
\end{figure*}

\subsection{Assets-based Community Cultural Wealth as a Framework}
\noindent Digital divide narratives often take a deficit-based approach, focusing on what resources that Black and Latine communities are deficient of, whether it be educational access, money, access to various forms of technology, or technology evangelism. The dominant mode of interpretation is that underrepresented minority communities are presumably lacking something that is preventing their agency within technology access and even representation in technical and engineering educational programs \cite{samuelson2016community, ovink2011more}. 

Social and cultural capital theories \cite{bourdieu1986handbook} have been used to examine factors in learning sciences and HCI design research~\cite{cunningham2022cost, wong2020needs} affecting adoption and engagement of communities with digital and information technology. In particular, when applying a deficit-based lens to Black and Latine student persistence and retention in predominately White STEM programs, the reality of institutional barriers, such as uneven educational funding, lack of faculty representation, and discrimination in college environments are often over-shadowed \cite{harper2010anti, grabsch2023using, parnell2023engineering}. In turn this inequitable framing localizes the problems and solutions to mitigate them within individual students or their communities~\cite{samuelson2016community}. 

Community-cultural wealth has foundational grounding in intersectionality and critical racial identity, which presents an assets-based lens to understanding the persistence of underrepresented minority students in engineering programs \cite{samuelson2016community, villalpando2005role}. Samuelson et al, (2016) examined various forms of cultural capital found to be used by students of colors in their families and communities: \footnote{Six original forms of cultural capital in community cultural wealth was outlined by Yosso (2005). Asterisk indicates those in which Samuelson, et. al (2016) defined for contextualizing minority engineering student experiences.} *aspirational, *navigational, social, linguistic, *familial, and *resistant capital \cite{yosso2005whose, samuelson2016community}. Through their study, they outlined four dominate types of capital in which African American and Latine students leverage different types of community cultural wealth and how these types of capital contributed to students' participation and persistence in engineering.


What is collectively recognized by these intersecting theories, is that an asset-based, rather than a deficit-based framing for understanding and broadening participation in STEM among an underrepresented minority group like Black and Latine students, would have much more focused orientation of how people are capable of using their identities and culture as tools for envisioning equitable access and transformative justice in STEM education. Community-based and student-led organizations like \avela{}, who empower Black and Latine students at a public predominately-White institution (PWI), are designed to engage in participatory efforts that promote accessible and justice-oriented educational programming. Within HCI, computing research, and engineering education, methods of community-based participatory design and research have recently emerged as practices that support equitable exploration and collaboration of underrepresented minority groups in technology design \cite{harrington2019deconstructing,cunningham2023collaboratively, parnell2023engineering}.

\subsection{Impact of Student-led Organizations}
Student-led organizations like NSBE and SHPE aim to prepare Black and Latine students for careers in STEM fields through student-led leadership, service learning, and mentor embodied community representation~\cite{nsbe-27280,nsbe-demographics,shpe-demographics}. 
Student organization has been shown to promote economic, social, and cultural capital amongst aspiring STEM scholars~\cite{nsbe-27280}. 
New student-led organizations like \avela{} leverage the infrastructure established by existing groups like NSBE and SHPE. From recruiting and marketing, to establishing budgets and university partners, to cost-sharing and student advocacy for campus activism, existing student groups have paved the way for growing organizations like the \avela{} to address old problems with new methodologies. 


\section{Positionality Statement}  

We are a team of researchers composed of 3 Black cisgender male PhD candidates, 1 Latino cisgender male PhD candidate, 1 Latina cisgender female master's student, 1  Black cisgender female undergraduate student, 1 Black cisgender female educator, 1 Asian cisgender male faculty advisor, and 1 White cisgender male advisor of Jewish decent. All student contributors are from lower to middle class backgrounds, and are the first in their families to attend college to study STEM. One of the Black male PhD students and the Hispanic male PhD student are the co-founders of \avela{}. As a team of majority Black and Latine students, we recognize the importance of STEM outreach programs and their effect on Black and Latine communities in urban areas. As Black and Latine students attending a predominately white institution, we have first-hand experience on how race, socioeconomic status, and family education intertwine to create an educational and technology access divide between our communities and the dominate communities. 

We recognize that our positionality affords us the ability to communicate with our participants on a personal level, however, our leadership positions within \avela{} may impact our research process and influence our interpretation of the data and conclusions we draw. Our positionality as a team of researchers with diverse social identities shapes our approach to this research and our interpretation of the data. By being transparent about our positionality, we hope to provide context for readers to better understand our findings. We focus our findings specifically on Black and Latine communities based on the lived experiences and expertise of our research team, community partners, and stakeholders.
\section{The \avela{} Model}

We describe the main components of \avela{}'s mentorship methods, leadership structure, and educational activities below. This model has been actively adapted over the years as the \avela{} continues to grow,
see Fig~\ref{fig:avela-timeline} for key milestones.

\noindent\textbf{Interactive \& Experiential Learning.}
\avela{} offers college students the opportunity to gain hands-on and research-based experiences developing and presenting STEM activities to secondary school students with the goal of bolstering the technical and leadership skills of Black and Latine students in both secondary schools and colleges.

\begin{figure}
  \centering
  \includegraphics[width=0.9\linewidth]{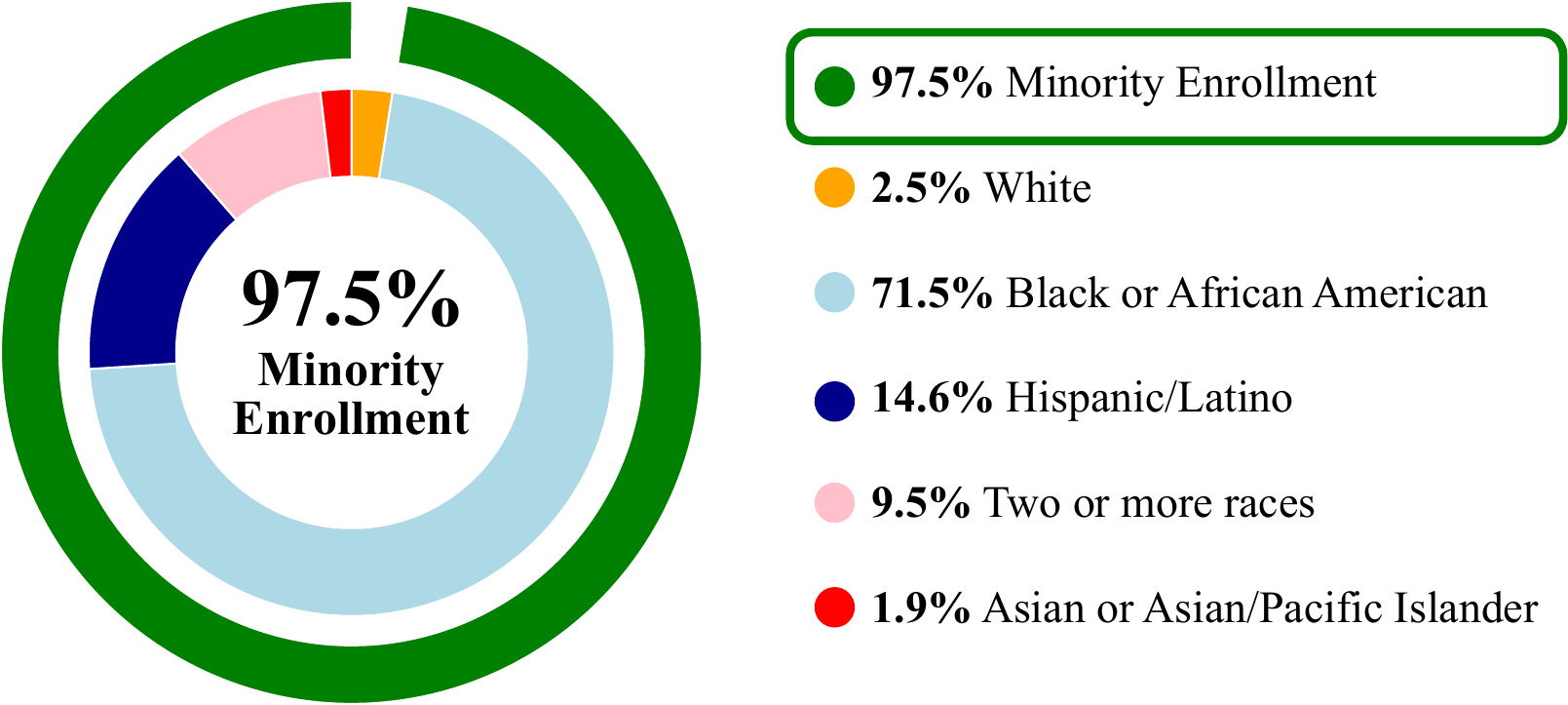}
  \caption{Demographic breakdown at one of \avela{}'s partnering schools.}
  \Description{Pie chart showing student demographics at a partnering school broken down by ethnicity. }
  \label{fig:ethnic-bacgkround-pie-chart}
\end{figure}

\noindent\textbf{Mentor Embodied Community Representation.}
\avela{} targets Black and Latine students (see Fig~\ref{fig:ethnic-bacgkround-pie-chart}) in communities that lack equitable access to STEM resources like physical hardware and technology, career mentorship from STEM professionals, college application assistance, as well as awareness of free online or local educational tools and resources. One example of this is at a local high school which in 2019, when \avela{} began partnering with them, did not offer any  computer science classes for their more than 800 students. \avela{} offered students 10-week Python and Arduino coding classes, which always filled to capacity with eager students. Similar to the students in \avela{}'s programming class, \avela{}'s university instructors had also not yet participated in a computer science class when they were in high school. This class, like other \avela{} classes, was taught by university students who shared identities with the high school students.

\noindent\textbf{Near-peer Mentorship.}
Undergraduate students in the \avela{} are mentored by graduate students,  professors, and industry allies, before they go on to mentor secondary school students interested in STEM. Through a near-peer mentorship model, the \avela{} provides Black and Latine students at all levels with culturally responsive and representative STEM experiences, as well as exposure to university-level and industry-standard tools and concepts. This kind of outreach and near-peer instruction has been shown to benefit both the secondary students and college student instructors~\cite{pluth2015collaboration}. 

\noindent\textbf{Student-led Community Engagement.}
The \avela{} began as a registered student organization at \uw{}, and is now also a federally recognized nonprofit. While \avela{} has similarities to NSBE's and SHPE's local chapters at universities, a key difference is that both levels of the \avela{} are entirely student-driven and led by members who are predominantly current students at or recent alumni from \uw{}. This model draws on recommendations from STEM persistence frameworks which emphasize student agency over institutional goals like retention, which have been demonstrated to increase student participation in STEM in both industrial and academic settings~\cite{graham2013}. \avela{} employs a persistence framework by focusing on student agency and near-peer mentoring, an approach that has yielded significant impact in terms of membership growth.

During the 2022-2023 academic year, more than 100 different \avela{} members from Black and Latine backgrounds participated in teaching an \avela{} class (37\% male and 63\% female or non-binary), with more than 200 members participating in events and other community building activities. \avela{}'s college instructors were able to teach to more than 1200 K-14 students in more than 50 classrooms and community centers across the state (91\% in-person and 9\% virtual/hybrid), with most activities concentrated in the \sea{}. 

\noindent\textbf{Culturally Responsive Lessons \& Academic Support.} 
The \avela{} model aligns with the large body of work indicating the importance of culturally responsive support groups~\cite{cr-pedagogy} and social belonging for the graduation and retention of Black and Latine students in college~\cite{hotchkins2019first, patterson2021you}. These evidence based practices~\cite{bergerson2014outreach, anderson2019benefits} have demonstrated concrete benefits to URM scholars through leadership and affirmation of their place in engineering. \avela{}'s outreach model has significant community impact outside of \uw{}, creating a combined solution to recruiting, retaining, and graduating Black and Latine scholars. Prior work has shown that there are benefits to same gender/race mentoring relationships ~\cite{syed2012individual}. Learning about diversity and inclusion also helps students develop a STEM identity and sense of belonging~\cite{singer2020}. Additionally, a sense of belonging has been shown to positively relate to program attendance and protective factors found in communities~\cite{belonging}.

\noindent\textbf{Compensating Student Mentors.}
Prior studies indicate that engaging in outreach helps develop a professional engineering identity, which is strongly linked to persistence~\cite{bergerson2014outreach,wright1987role}. However, college students have to stay on top of their academics, family obligations, health, extracurricular activities, jobs and other finances. Finances being critical for the disproportionately low-income Black and Latine student mentors, who often work campus jobs to support themselves~\cite{pell_urm_all_UW}. We find that representative outreach is only sustainable and scalable when compensating student instructors for their time, enabling them to create and lead STEM outreach courses directly related to their course of study in place of nontechnical campus jobs. 
Compensating students is key in recruiting representative instructors, and empowering them to focus on leading quality STEM classes by alleviating the financial burden of college while building relevant technical and leadership skills.

\noindent\textbf{Co-instruction \& Class Enrollment Thresholds.}
Each \avela{} class requires at least two instructors to be present. Younger and less experienced members are paired with more experienced instructors to help them in learning how to teach STEM classes. This practice is implemented to increase the reliability of student-led classes. Instructors work collaboratively to create the teaching content and prepare any ancillary lesson materials. 

Low-income students have been shown to benefit more from smaller class sizes compared to students who are not low-income \cite{krueger2001effect}. \avela{} works in predominantly low-income areas, therefore we limit the enrollment to 15 students per instructor. The duration and time commitment of each class is determined in collaboration with the teaching staff and students at each school, as well as with the \avela{} instructors. 

\noindent\textbf{Social Good \& Research-driven Lessons.}
The \avela{}'s synergy with local communities and universities has led to the development of hundreds of project-based lessons, many focused on various aspects of STEM for a social good or the \avela{} members' research projects. Research-based lesson plans draw on the expertise of student instructors and exposes secondary school students to emerging technologies, demonstrating role models and career pathways. The \avela{} has also been able to help faculty apply for NSF REU awards, which has helped fund paid research experiences for 8 different members from the \avela{} last year alone. The \avela{} has also partnered with the Society of Professional Engineering Employees in Aerospace union and other professional labor organizations. Through these partnerships, \avela{} helps host professional development and career exploration classes, as well as other lessons tailored to help students solve problems in their own communities. A handful of these projects are described below, and illustrated in Fig~\ref{fig:avela-classes}:

\begin{figure*}
  \centering
  \includegraphics[width=1.0\linewidth]{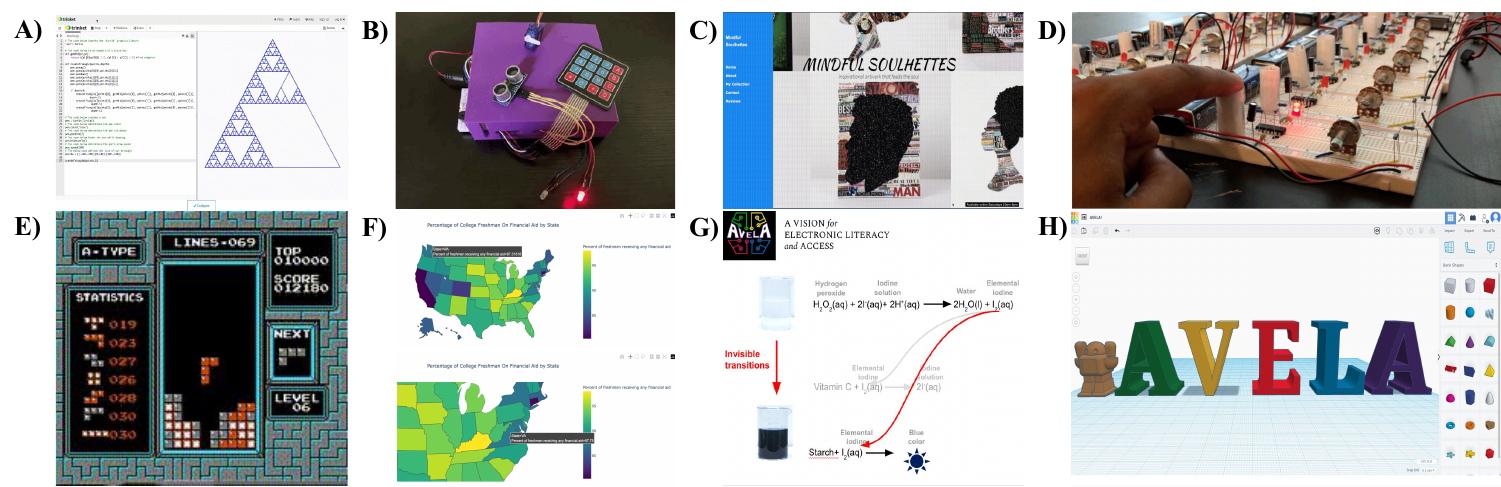}
  \caption{Various \avela{} projects are shown with their names listed as follows, with more detailed information on the projects specified in Section 4 Social Good \& Research-driven Lessons. \textbf{A)} Coding fractals using Python. \textbf{B)} Portable door alarm system using Arduino. \textbf{C)} Website development for minority owned businesses using HTML and CSS. \textbf{D)} Analog heart rate monitor circuit costing only \$10. \textbf{E)} Machine learning for Tetris player artificial intelligence (AI). \textbf{F)} Data activism using Python coding libraries. \textbf{G)} Iodine clock using measured chemical reaction times. \textbf{H)} Sustainable design using biodegradable 3D filament.}
  \Description{Figure showcasing eight different class projects created during an \avela{}al workshop. Figure A is a screenshot of a coded Sierpinski triangle. Figure B is a photo of the lock system alarm setup. Figure C is a screenshot of a website created by \avela{} students. Figure D is a photo of the analog heart rate monitor setup. Figure E is a screenshot of the tetris game an \avela{} student coded in a class. Figure F is a heatmap of the United States that an \avela{}al student coded using python. Figure G is a screenshot of a lesson plan including an Iodine clock with it's corresponding chemical equation. Figure H is a CAD model of a sustainable design using biodegradable 3D filament workshop.}
  \label{fig:avela-classes}
\end{figure*}


\smallskip\noindent\textit{Coding Fractals!} Students handcraft fractals, beautiful self-repeating geometric patterns, out of paper before visualizing various fractals using an online Python IDE. Students learn how software can be a powerful tool for visualizing geometric patterns that are tedious to make by hand.

\smallskip\noindent\textit{Portable Door Alarm System.} Students learn about basic circuit theory through a series of Arduino mini-projects, culminating in the Portable Door Alarm System final project. Each mini-project focuses on reading or showing data from circuit components like LEDs, servos, keypads, distance sensors, and more. These Arduino sensors are used to accomplish tasks like ringing a buzzer and making a servo-based timer. 

\smallskip\noindent\textit{Website Development for Local Minority Owned Businesses.} Students work with local minority owned businesses to design and create professional websites in alignment with the vision of the partnering company. Students learn how to code these websites using HTML, CSS, and Javascript.

\smallskip\noindent\textit{Analog Heart Rate Monitor.} Students learn electrical engineering fundamentals like Ohm’s Law, Kirchoff’s Voltage and Current Laws, as well as filter design with operational amplifiers. With this knowledge, students are able to simulate a working design of an analog heart rate monitor on LTspice before getting to build their own using purely analog components.

\smallskip\noindent\textit{Machine Learning a Smart Tetris AI.} Students first get exposed to basic coding concepts before learning about the machine learning, a form of Artificial Intelligence (AI), concept of state machines. Students reflect on their own learning behavior while playing Tetris, before learning how to code an AI agent with optimized Tetris playing behavior to beat their personal scores.

\smallskip\noindent\textit{Data Activism.} Students learn how technology and data can be a tool for activism. After learning basic Python coding concepts, students are tasked to use various data management libraries to create graphics like pie charts, heat maps, and other visualizations using publically available college admissions data at HBCUs (Historically Black Colleges and Universities) and PWIs.

\smallskip\noindent\textit{Iodine clock.} Using known and measured chemical reaction times, students are exposed to various reactions before making their own iodine-based clock. This exposes students to research concepts like reaction kinetics, non-linear chemical systems, numerical modeling, molecular self-assembly, and biological cycles. 

\smallskip\noindent\textit{Tinker aCADemy - Sustainable Design using 3D Modeling \& Printing.}
Through the tedious creation of handmade 3D shapes out of paper students will get to feel for themselves the motivation behind using Computer Aided Design (CAD) software to rapidly prototype 3D structures. Students are then tasked with using CAD software to design their own creation given a combination of structural, environmental~\cite{tale}, and cost constraints. 
\section{Methods}
In this study we conduct semi-structured interviews with student instructors from \avela{}, gather survey responses from secondary school students who participated in \avela{}'s classes, and review documents and other artifacts used by the leadership of the \avela{} over the past 4 years. We interviewed 24 student participants in \avela{} and gathered survey data from a pre-interview survey study, followed by a survey study on 171 \avela{} class participants during the 2022-2023 academic school year. Insight on these data is provided through an autoethnographic analysis.

The complete lists of interview, pre-interview survey, and class feedback survey questions are provided in Appendices A, B, and C respectively. The study protocol was approved by our Institutional Review Board (IRB) and we obtained informed consent from all interview and survey participants, as detailed in Appendix section B.

\subsection{Semi-structured Interview Study}
We recruited participants by sending out study invitations via email to students who had prior experience teaching a class or workshop through \avela{}. Interview participants completed a pre-interview survey, which assessed their involvement in \avela{}, their access to technology and education, as well as their demographic background (see Appendix section A). Once the screening was complete, students were able to schedule an interview. All interviews were conducted virtually over Zoom and were audio-recorded. We used a combination of Zoom and OpenAI's Whisper for transcription. Each session lasted approximately 30 to 45 minutes, and each participant received \$20 in compensation for taking part in the study.

\begin{figure*}
  \centering
  \includegraphics[width=0.8\linewidth]{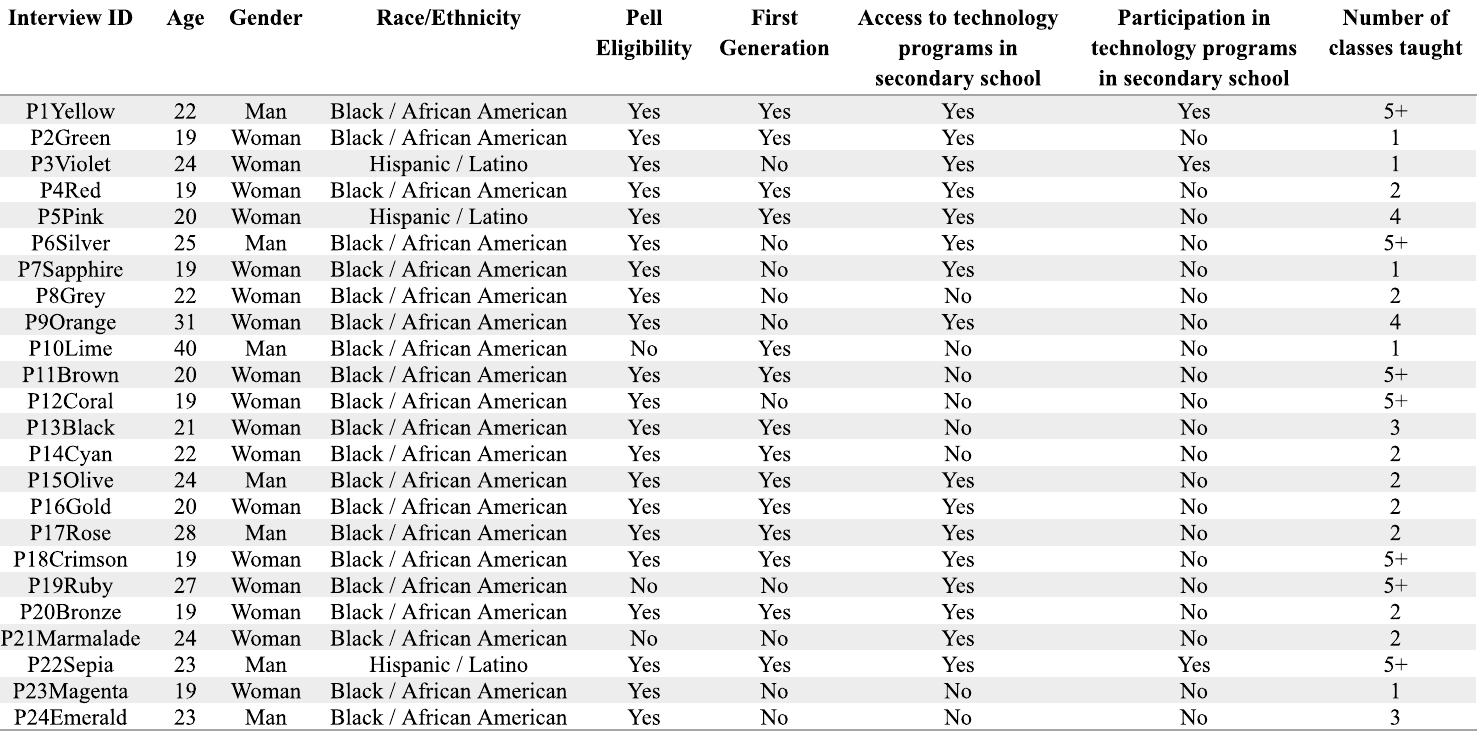}
  \caption{Overview of interview participants’ demographic information.}
  \Description{Overview of interview participants’ demographic information. Majority of participants were Pell Grant eligible, first-generation college students, had some sort of access to technology in secondary school, and did not participate in said programs in secondary school. The complete demographic information of the interviewees can be found in the Appendix.}
  \label{fig:qual-demographics}
\end{figure*}

\subsection{Student Feedback Surveys (pre- and post-class)}
Student feedback was collected from secondary school students using pre- and post-class surveys with a 5-point Likert scale. The surveys assessed students' prior knowledge and experiences before taking classes (pre-class survey) as well as after completing the classes (post-class survey). The survey included questions that assessed students' comfort in their current STEM knowledge, prior knowledge of STEM subjects, educational and career aspirations as they relate to STEM fields, and their belief on whether STEM can have a positive impact on their community. Additionally, participants were asked about the influence of STEM role models in their lives. Both surveys were completed online using Google Forms. 

\subsection{Autoethnography}
Each quarter of operation, \avela{}'s founders meet with the student organization's board to reflect on student impact and success based on the class feedback surveys and concerns shared with the \avela{} board viva voce. Leaders on \avela{}'s board also meet monthly with the \uw{} engineering department's Associate Director for DEI to share resources and strategies on conducting outreach. An abundance of feedback from corporations, school districts, principles, and community leaders was also shared at \avela{}'s monthly nonprofit meetings. A subset of the authors of this paper were the founders of \avela{} and participated in these meetings.

\subsection{Data Analysis}
We conducted qualitative data analysis to examine the interview responses of participants. Additionally, we applied statistical methods to quantitatively analyze participants' pre and post-class survey responses. We also used a combination of thematic, textual, and visual analysis on \avela{} artifacts.

\subsubsection{Qualitative Analysis}
We conducted 24 semi-structured interviews with Black and Latine \avela{} student instructors who taught a STEM class or workshop. We used a mix of inductive and deductive coding to analyze data and explore the perceived effectiveness of \avela{}'s student-led STEM engagement model. Initially, we reviewed all interview transcripts thoroughly and then used a deductive approach to establish a list of 27 codes based on the research and interview questions. In the next phase, two of our study's researchers began applying the list of codes we originally created to the relevant text, while also creating 16 new codes as we saw new patterns emerge. We utilized Dedoose to develop codes, code the transcripts, and develop themes. After the completion of our final codebook (see Appendix section D), the two study researchers went through each of the excerpts to make sure they met the code definitions. For the thematic analysis phase, we systematically reviewed all the codes within Dedoose, employing features like code co-occurrence and application analysis to identify patterns and redundancies. Through this process, we identified a total of three key themes.

\subsubsection{Quantitative Analysis}
Quantitative data was collected through pre- and post-class surveys administered to secondary school students. Most questions in these surveys utilized a 5-point Likert scale with the options: strongly agree, agree, neither agree nor disagree, disagree, and strongly disagree. On the first day of class, students were given the option to fill out a pre-class survey to assess their prior knowledge and experience with the courses' STEM topics. The students were given unique interviewee identification (ID) pseudonyms for confidentiality. Both pre- and post-class surveys were completed online using a standard survey format via Google Forms. We utilize an unequal variance t-test analysis to compare the pre- and post-class survey results, and further derived a Cohen's d effect size. 
%
%

\subsubsection{Autoethnographic Analysis}
The main method used for our autoethnographic study was the collection and analysis of personal artifacts. This included gathering \avela{} meeting notes, components from past \avela{} projects, digital materials, and visual media forms like flyers and posters. As a research team we met to reflect on our experiences in \avela{} to help us provide deeper insights into our experiences as Black and Latine students in STEM. Some artifacts that were examined include Arduino circuits, a prototype analog heart rate monitor breadboard, the initial design for the first \avela{} logo, digital folders containing \avela{} class photographs and videos, lesson plans and class presentation slides, social media posts, emails, and other STEM activity materials utilized throughout various \avela{} classes. These artifacts provided rich, concrete prompts for reflection and interpretative analysis. We cataloged and organized artifacts chronologically and thematically. For each item, we shared a memory describing its significance in representing aspects of \avela{}'s enculturation. We then looked for patterns in how the artifacts evidenced cultural values, struggles, milestones, and pain points. The goal was to leverage the artifacts as data to gain an in-depth understanding of our own acculturation process. This method provided a tangible, evocative avenue for systematically examining the interplay between our individual biographies and the culture of \avela{}.
\section{Results}

A total of 24 college students participated in the qualitative interview, see Fig~\ref{fig:qual-demographics} for the breakdown of characteristics of each \avela{} student participant. In summary, the average age of respondents was 22.9 years old, 70.8\% identified as women, 87.5\% were of a Black/African American background, 12.5\% were of Hispanic/Latine background, and 87.5\% were Pell-eligible (an indicator of coming from a low-income family). Additionally, 58.3\% were first-generation college students, 66.7\% had access to STEM programs in secondary school, and 12.5\% participated in technology programs in secondary school. In total the study participants' taught more than 70 \avela{} classes across the \wa{}.

\begin{figure}
  \centering
  \includegraphics[width=1.0\linewidth]{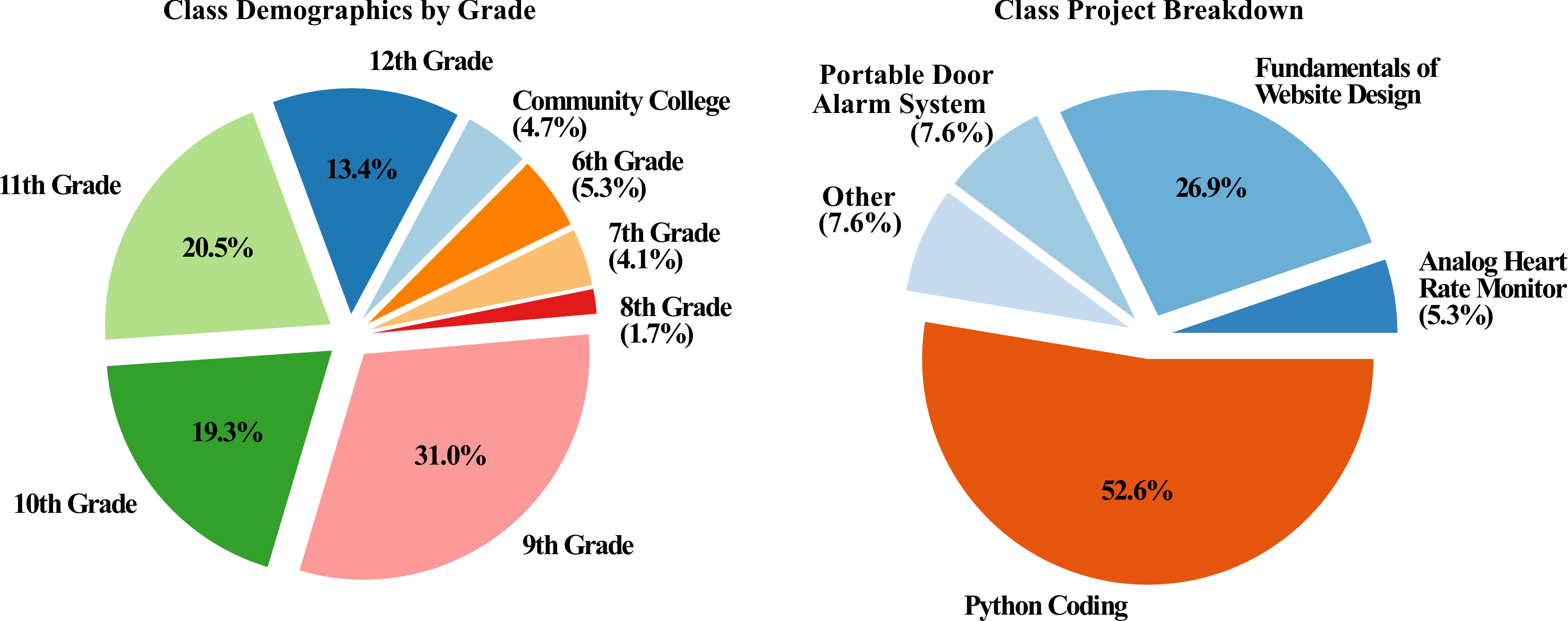}
  \caption{Class survey respondents’ demographics by grade (8 different grades) and breakdown of class projects (9 different projects). The complete demographic information of the interviewees can be found in the Appendix.}
  \Description{Two separate pie charts giving an overview of survey respondents’ demographic information. The complete demographic information of the interviewees can be found in the Appendix.}
  \label{fig:quant-demographics}
\end{figure}

A total of 171 secondary school students completed the pre- and post-class survey (see Fig~\ref{fig:quant-demographics} for \avela{} class participant characteristic breakdown). In summary, 11.1\% of the students were in middle school (grades 6-8), 84.2\% were in high school (grades 9-12), and 4.7\% were in community college. Of those surveyed, 52.6\% of respondents participated in a Python coding class (Coding Fractals!, Machine Learning a Smart Tetris AI, Data Activism, etc.), 26.9\% in the Fundamentals of Website Design class (for Local Minority Owned Businesses), 7.6\% in the Portable Door Alarm System class (introduction to Arduino), 5.3\% in the Analog Heart Rate Monitor Class, and 7.6\% in other \avela{} STEM classes. 

Below, we describe the following three key themes that emerged from our multimethod analysis on sociotechnical access inequalities for Black and Latine urban communities and the impact of the \avela{} student-led STEM engagement model at mitigating the digital divide:
\begin{itemize}
\item\textbf{Problem.} Hesitance to engage with available technologies due to lack of support or know-how.

\item\textbf{Approach.} The importance of representative and culturally responsive mentorship.

\item\textbf{Outcome.} Skill development and career enhancement through service learning.
\end{itemize}

\subsection{Theme 1: Hesitance to engage with available technologies due to lack of support or know-how}

Most of the students reported a reluctance to engage with the technologies at their disposal. Despite being equipped with resources such as school-issued laptops, technology-related classes, access to school or library Wi-Fi, and various other technological tools, students frequently expressed their hesitancy to actively utilize these resources. The underlying factor contributing to this hesitation appeared to be a lack of understanding about how to effectively interact with these technologies. For example, P1Yellow said: 
\begin{quote}
"\textit{The biggest barrier was not understanding it (technology) because I didn't have much access to it when I was younger}" [P1Yellow]. 
\end{quote}
Many of the students found themselves in possession of laptops, but these devices primarily served as tools for basic assignments and word processing. Beyond these elementary functions, the broader potential and exploratory possibilities of these devices remained largely undiscovered due to the students' uncertainty about what more they could achieve. Several students mentioned having to navigate learning about new technologies on their own, P14Cyan reporting: 
\begin{quote}
"\textit{Beyond like just, you know, going on YouTube or doing assignments, there wasn't much that I knew about technology. I didn't know, like how I could, I guess, use the computer that I had as a tool to sort of learn more}" [P14Cyan].
\end{quote}
These quotes underscore the importance of addressing knowledge gaps and providing technical support to students so that they are less hesitant to engage with available technology. In the pre-class survey, before any STEM intervention, students reported an average desire to pursue a STEM career at 3.52 +/- SD=1.23 (1-5 scale). In the same pre-class survey, students rated their belief of having a positive impact on their communities at 3.54 +/- SD=1.12 (1-5 scale). In the pre- and post-class surveys, 78\% of students held aspirations towards STEM careers and 50\% held the belief that they can make positive contributions to their community through STEM. 

These statistics underscore a genuine interest from students in wanting to participate in technology-related fields, an interest that we can also validate from personal experiences. Upon further reflection, we acknowledge that most students who join \avela{} do so out of a curiosity about STEM and an intrinsic desire to support their own households and neighborhoods. Many \avela{} students have shared with the founders that they were 'sold' on joining \avela{} after hearing how \avela{} uses STEM for a social good, focusing on projects around community-centered designs and impact. The importance many Black and Latine students place on social good and impact may explain why a student who only uses a computer to type out a homework assignment would not see a reason to further engage with the technology. This reinforces the qualitative theme that highlights Black and Latine students' hesitance to participate in STEM originating from a lack of know-how on how to use the devices physically available to them. 

Moreover, this hesitation to engage with devices is further exacerbated by a perceived absence of support from school staff and adults. P13Black reported:
\begin{quote}
\textit{“Like my parents don't really speak English, and they don't know how to use like that technology. So I had to figure it out myself"} [P13Black]. 
\end{quote}
P10Lime shared: 
\begin{quote}
\textit{"I was part of a group in my generation that had access to computers. Maybe the first few people who had access to computers were coming into the curriculum and becoming common. And we didn't have people who were educated to help us make the most out of the gadgets themselves. One, the gadgets weren't very good. They were definitely second-hand gadgets that were just being donated in our communities. So, there was also no one who was good at teaching about them because nobody had been trained to use them"} [P10Lime]. 
\end{quote}

\begin{figure}
  \centering
  \includegraphics[width=1.0\linewidth]{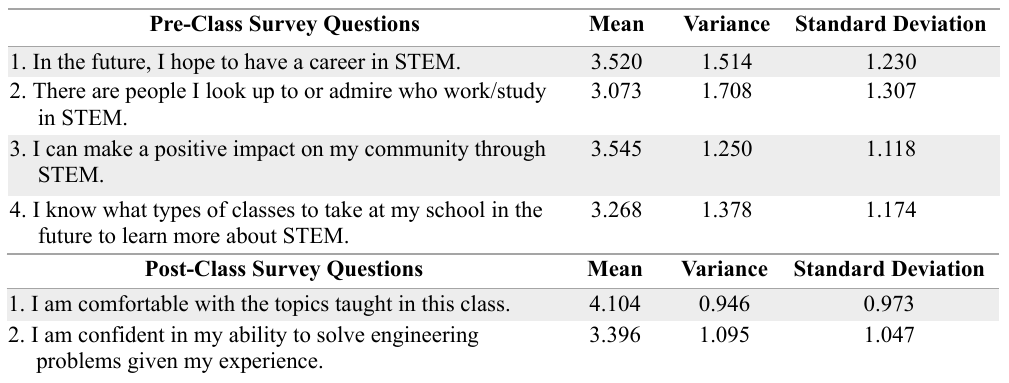}
  \caption{Summary statistics of student survey responses regarding STEM interest and confidence before and after \avela{} classes.}
  \Description{Pre- and post- survey statistics. The table displays the mean, variance, and standard deviation for student responses to survey questions before and after participating in \avela{} classes.}
  \label{fig:survey-stats}
\end{figure}

\begin{figure}
  \centering
  \includegraphics[width=1\linewidth]{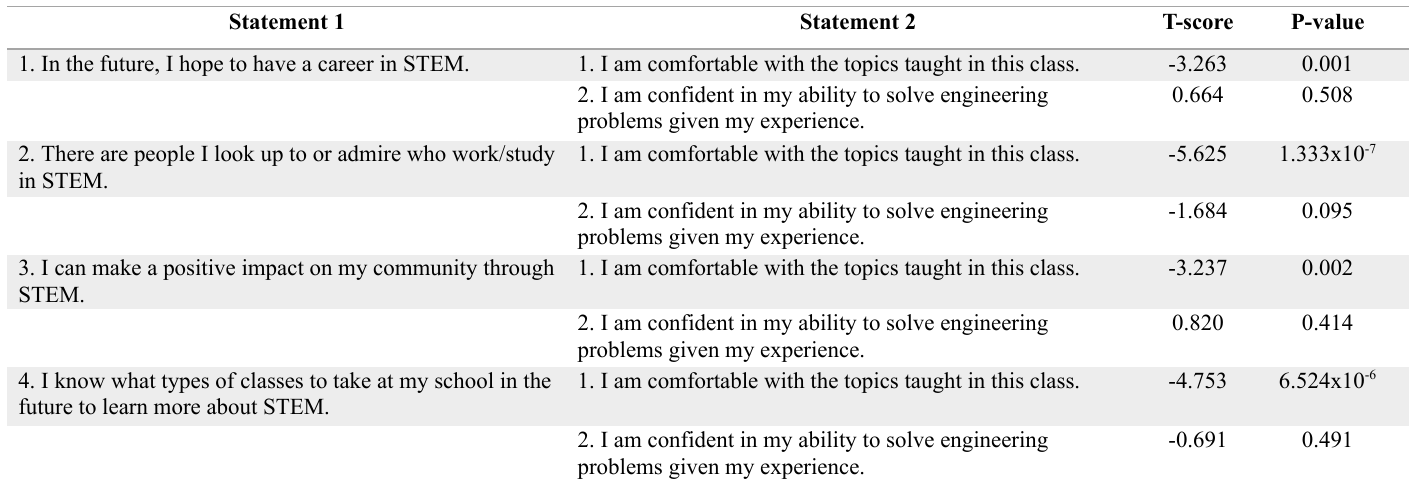}
  \caption{Summary statistics of t-scores and p-values for pre- and post-class survey questions.}
  \Description{This table shows the results of t-tests and p-values comparing pre-survey responses to the two post-survey responses.}
  \label{fig:comparison-stats}
\end{figure}

From our \avela{} class survey data, more than 50\% of students across all grade levels disclosed having minimal to no prior knowledge of STEM topics before participating in an \avela{} class. Having limited exposure to technology during one's formative years may offer one explanation to why many students may feel unprepared to interact effectively with technology in secondary school. Feeling unprepared for STEM due to having limited background knowledge on the subject also showed to affect students' belief in their ability to have a positive impact on their community. By comparing responses from the pre- and post-class survey, we see that students' belief in making a positive impact on their communities through STEM and their comfort with the STEM topics taught in the \avela{ } classes yields a medium effect size (Cohen's d=0.49). This suggests a medium positive effect that the \avela{}'s approach to providing technical support and exposure to STEM concepts enhanced student confidence in making a positive impact in their communities through STEM (see Fig~\ref{fig:survey-stats} and Fig~\ref{fig:comparison-stats}), while highlighting the need to provide more STEM support to students in order to further alleviate hesitance towards technology engagement. This reinforces the main theme that highlights Black and Latine students' hesitance to participate in STEM originating from a lack of technical support and know-how on how to use the technologies that are physically available to them. 

\subsection{Theme 2: The importance of representative and culturally responsive mentorship}

Students in \avela{} consistently emphasized the value of having role models who both represented diverse cultural backgrounds and were also engaged in technology-centered careers. These mentors served as examples of inspiration, offering relatable stories of success and guidance that resonated deeply with the students. This highlighted the importance of co-instructing, specifically with pairs of more experienced and less experienced teachers. P7Sapphire shared the impact of receiving mentorship within \avela{}: 
\begin{quote}
\textit{“What I liked was I felt [mentor name] had a lot of wisdom about, you know, being a person of color in the STEM field and [they] had a lot of perspective and a lot to share in a lot of experiences that like opened my eyes to what it can be like. So that was super helpful”} [P7Sapphire].
\end{quote}
P10Lime emphasized how their work with \avela{} had an impact on others: 
\begin{quote}
\textit{"My work with \avela{} was kind of an aha moment to just reinforce what I've done over the years and why I keep on doing those things and working with certain communities and what are the common threads around them. And it's technology, it's mentorship, it's access, it's, you know, just making people of color feel like there is a space for them and, you know, giving them that strong foundation to build upon and just, you know, perhaps, you know, taking them to a point where you can say these are the possibilities and connecting them to careers, connecting them to networks and opening up opportunities for them"} [P10Lime].
\end{quote}
Additionally, compensation was a significant factor in the program. As P4Red mentioned:

\begin{quote}
\textit{"Money definitely helped because I've kind of struggled with money a little bit. So it was kind of like a way for me to do a job and get paid for it..."} [P4Red].
\end{quote}

The mentorship provided by \avela{} extended beyond inspiration; it actively linked students with opportunities and established the groundwork for a representative and inclusive technology landscape. In our interviews with 24 \avela{} student instructors, 87.5\% of these mentors came from Black/African American backgrounds and 12.5\% were of Hispanic/Latino backgrounds. Additionally, 70.8\% identified as women, spotlighting \avela{}'s commitment to both ethnoracial and gender diversity in STEM. We also discovered that these \avela{} mentors collectively spoke 12 different languages, highlighting the importance of diverse representation even amongst the same ethnocultural group. This data reminded the authors of a coding class they taught at a local community center with predominantly Black students (80\% Black), where an \avela{} instructor discovered that one of the students participating in the workshop did not speak English and was thus struggling to follow along with instructions. During the next lesson, the instructor was able to bring another \avela{} member along with them to help translate the instructions for the student. This anecdote reiterates the significance of representative and culturally aware mentorship for Black and Latine communities. 

Our autoethnographic analysis exposed another key insight into the importance of mentors who can relate to the experiences and cultures of their students. A few months after teaching a coding class at a local charter school, one of the \avela{} founders was informed that a student who didn't initially apply to computer science (CS) had called the admissions office and asked to switch their majors to CS. The Office of Admissions allowed the student to switch majors, and they were subsequently admitted into the CS major the following term. This student cited their participation in an \avela{} coding class, and specifically their interactions with the \avela{} founder who represented their same background, as the driving factor behind wanting to switch majors. 

This experience reminded the founders of the importance of representative mentorship, and the impact it can have on a student's comfort level or acceptance in STEM spaces. When comparing responses from the pre-class survey on students' who have someone with whom they look up to in STEM with responses from the post-class survey on their comfortability with STEM topics, we see significant change in attitudes (Cohen's d=0.86, a large effect size). Furthermore, 73\% of students expressed a strong interest in delving deeper into the subjects they learned in their \avela{} class. This data suggest that exposure to representative mentorship has a measurable impact on a students' comfortability with STEM topics, reinforcing the importance of creating an environment where students from diverse backgrounds can thrive, find inspiration, and pursue meaningful growth and impact leveraging representative and culturally responsive mentorship.

\subsection{Theme 3: Skill development and career enhancement through service learning}

\begin{figure*}[t]
  \centering
  \includegraphics[width=1.0\linewidth]{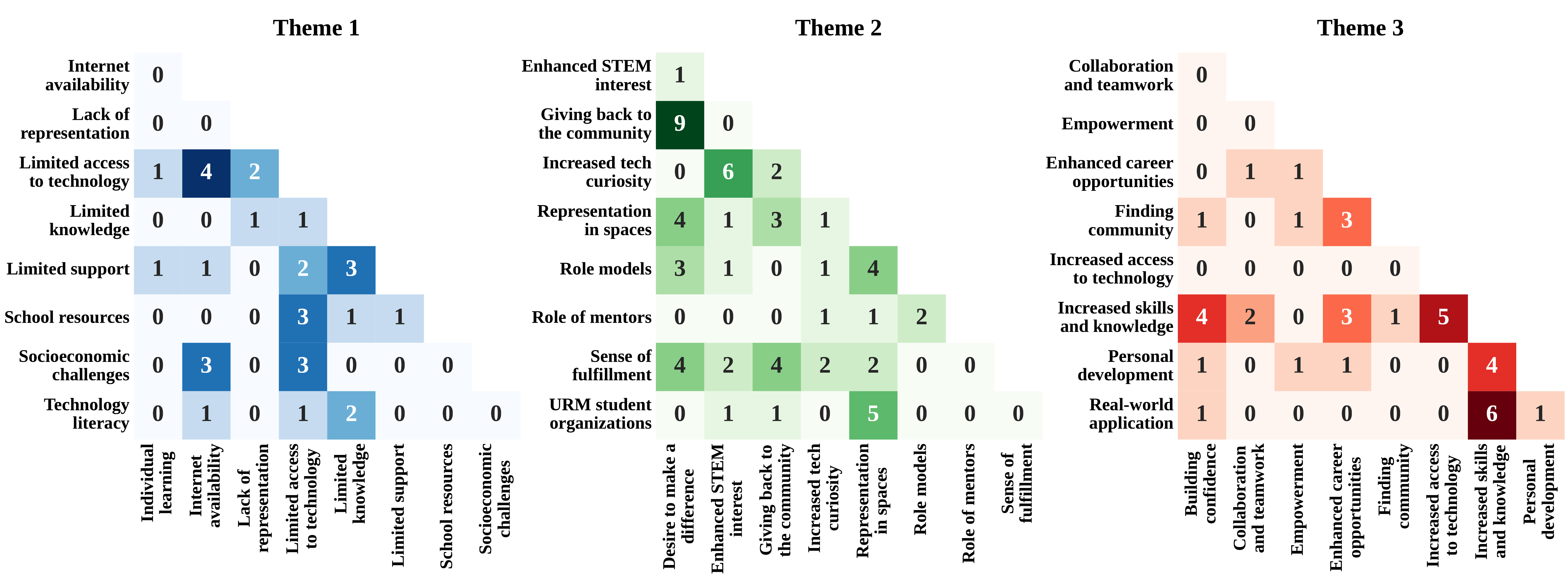}
  \caption{Co-occurrence counts for the three themes highlighted by this work on STEM education and accessibility.}
  \Description{Three separate charts for each theme with counts for co-occurrence of issues/themes presented during interviews, with the highest counts being shaded in a darker color. }
  \label{fig:co-occurrences}
\end{figure*}

Skill development, real-world applications, and career enhancement were prominent outcomes among student instructors. Students articulated the impact of their engagement with \avela{}, revealing how \avela{}'s initiatives served as a catalyst for them in cultivating a broad range of STEM competencies (see Fig~\ref{fig:co-occurrences}). These newfound skills were not merely confined within \avela{} settings, but were actively applied in other practical contexts as well; empowering students to confront diverse challenges and seize new opportunities. For example, P11Brown stated: 
\begin{quote}
\textit{ "I was able to learn how to, like professional development, for example, I think \avela{} has done a really good job in like emphasizing how to be professional and how to communicate and network. And so I think those skills are definitely skills that I use on a day to day basis"} [P11Brown].
\end{quote}
Most notably, \avela{} interventions translated into a tangible enhancement of career opportunities, as students found themselves better equipped and more confident in pursuing technology-based career pathways. P20Bronze shared: 
\begin{quote}
\textit{"I went to \uw{} originally for medicine, and then I went to the NSBE Conference, and I figured out I didn't want to do medicine anymore, and I want to do something tech. And then, after talking to people in \avela{} that kind of like solidified what I want to do in tech, which is like really focusing on like equitable tech"} [P20Bronze].
\end{quote}
P24Emerald discussed how \avela{} created pathways to new opportunities for them: 
\begin{quote}
\textit{"I thought I was going to go into academia before I started at \uw{}, so I thought... I do a master's, a Ph.D., etc... but \avela{} in a sort of way brought me to the real world and was part of my catalyzing factors on how the world is what we make of it, and with \avela{} specifically, I'm playing with three fields. I'm playing with tech, I'm playing with business, and I'm playing with entrepreneurship and because of that, it's made me open my options and think, OK, I have a degree in math and I can play with it however I want to..."} [P24Emerald]. 
\end{quote}

These quotes highlight the dynamic role played by \avela{} in nurturing not only the skills, but also the aspirations of Black and Latine students, bridging the digital divide by creating a pathway toward fulfilling and meaningful technology-based careers. Before participating in \avela{} classes, a notable 67\% of students reported having negligible knowledge about the subject matter. Post-instruction data reveals a remarkable transformation, with 96\% of students expressing newfound confidence in their STEM skills. Post-class survey responses also show that students are significantly more likely to enroll in additional STEM classes after participating in an \avela{} class. 

We observed a medium effect size when comparing responses from the pre-class survey on students' knowledge of STEM classes that they can register for at their schools with responses from the post-class survey on their comfortability with STEM topics (Cohen's d=0.73), with 73\% of students expressing eagerness to delve deeper into the STEM subjects they were introduced to. This suggests that \avela{}'s service learning initiatives have not only improved skill development and career enhancement, but also sparked a newfound interest and motivation among students to engage in technology-related fields. One \avela{} member even shared with the two \avela{} founders that they had emailed every professor in their department asking for an opportunity to do research, with no one responding after 3 months. Through an \avela{} partnership with a professor hosting a National Science Foundation (NSF) Research Experience for Undergraduates (REU), this student was connected with a summer research experience within a month of joining \avela{}. The unique combination of practical service learning experiences that allow for leaderships roles in teaching and exposure to academic research, professional career development, as well as STEM skill building activities has led \avela{}'s members to jobs at Microsoft, NASA, Amazon, Boeing, IBM, Texas Instruments, and Nvidia, as well as graduate programs at \uw{}, UCLA, UC Davis, UC Berkeley, and Johns Hopkins. These data strongly suggest that through \avela{}'s student-led STEM engagement model members are given unique access to opportunities that provide skill development and career enhancement through their service learning.
\section{Discussion}

This work presents and analyzes a new design experiment framework for empowering student-led STEM engagement in Black and Latine urban communities. The results of this multimethod study highlight the nuanced sociotechnical access gaps that persist for Black and Latine students from urban communities, even with increased physical access to devices and connectivity. While infrastructure barriers have decreased, systemic obstacles around identity, belonging, support, and empowerment endure. Our interviews uncovered strong hesitation among students to engage with available technologies, rooted in their limited technical know-how and inadequate technology support. 

Critically, this study underscores the vital role of a many-methods approach to building approachability and identity safety for Black and Latine students in STEM. Our data showed significant increases in students' comfort and interest in STEM after being exposed through \avela{}'s student-led STEM engagement classes. 
This affirms the \avela{}'s many-methods approach, which incorporates near-peer mentorship, interactive \& experiential learning, social good and research-driven lessons, culturally responsive academic support, mentor embodied community representation, and student-driven community engagement.

Unlike previous frameworks that focus on a single or handful of intervention strategies, we demonstrate the value of a many-methods approach that combines a multitude of pedagogical practices tailored to the needs of Black and Latine students. Unlike prior works that examined individual techniques like service learning or culturally responsive teaching, our findings reveal that fusing multiple evidence-based practices creates a holistic model for identity safety and STEM engagement. We empirically evaluate the impact of the \avela{} model, and grounds our results in the lived experiences of students through participatory design. All of our quantitative, qualitative, and autoethnographic results affirm that a multiplicity of mutually reinforcing methods can be greater than the sum of its parts to provide a more holistic solution to the multifaceted needs of Black and Latine students.


Survey and interview responses demonstrate URM student groups like NSBE, SHPE and \avela{} help students feel represented in STEM spaces. Students correlated these representative spaces to an increase in STEM role models and a desire to give back to their community (see Fig~\ref{fig:co-occurrences}). Students reported that giving back to their communities also increased their sense of fulfillment in STEM. Students expressed new confidence in their technical abilities, and eagerness to pursue future STEM engagements after participating in classes with real world applications. Increased skills, knowledge, and confidence were also associated with enhanced career opportunities and access to technology. Students noted that top-down STEM programming centered around stereotypical ‘pipeline’ careers can seem alienating or extractive. 
Side by side learning and near-peer mentorship have been shown to reduce these negative feelings~\cite{disruptive-teaching,anderson2019benefits}, especially when implemented through student-led initiatives~\cite{nsbe-27280}.

\subsection{Barriers and Challenges}
We highlight the following challenges and provide recommendations for academic stakeholders to consider when conducting outreach to Black and Latine communities.

\noindent\textbf{Representation.}
The limited pool of college students from Black and Latine backgrounds constrains \avela{}'s recruiting. With only around 3\% of the \uw{} student population identifying as Black/African American, it is challenging to find enough mentors to meet the broader city and state demand. This bottleneck around representative near-peer mentors, critical to \avela{}'s approach, may restrict this approach from expanding. We hope students taking \avela{}'s classes continue to expand this recruiting base, as we have already seen our approach establish a positive feedback loop for continued growth. 

The underrepresentation of Black and Latine professionals in higher education and tech industries also means that there are fewer potential corporate partners, advocates, and role models willing to support \avela{}'s mission. Mainstream STEM programs and resource providers often harbor cultural disconnects from the communities \avela{} aims to serve. The lack of diversity in these ecosystems places strain on \avela{} as a student-led nonprofit navigating spaces not designed for their model and consitituencies. Until dominant norms, incentives, and power structures shift to more openly embrace diversity and community-centric frameworks, grassroots groups like \avela{} may remain limited in the scale of their models impact.

However, the very existence of \avela{} highlights the resilient change that marginalized youth can drive even in inequitable systems. The motivation and ingenuity within Black and Latine communities cannot be understated. Similar to the growth of NSBE and SHPE, \avela{}'s goal is to push institutions to reimagine access, achievement, and leadership in STEM by proving our communities are capable and interested in participating.

\noindent\textbf{Sustainability of Student Initiative.}
The \avela{} student-run model risks fizzling out as core members advance in their careers requiring new cohorts to rebuild lost expertise, connections to faculty and administrative partners, university resources, and STEM materials and tools. To avoid this, \avela{} strives to maintain the financial capacity to hire full-time staff. Leveraging a paid mentorship model should help continue to attract new students to the organization, with full-time staff helping maintain operational continuity and mentor training. Persistent institutional infrastructure to maintain grant funding, partnerships, and non-profit status are critical for the \avela{} model to scale despite student turnover. 




Moreover, \avela{} strives to increase sustainability through several strategies. Alumni are engaged as mentors and advisory board members to retain knowledge. Shared virtual drives store documents and assets to avoid reinventing resources. Subcommittees focused on fundraising and strategic planning aim to strengthen organizational backbone. Still, threats related to scale, consistency and resourcing persist due to the inherent challenges of a student-led model operating within the university context. Sustainability will demand ongoing creativity, agility and institutional collaboration. Ultimately, \avela{}'s goal is to demonstrate the capabilities of students to innovate new frameworks that don't yet exist, and sustainability may just be one more system needing to transform.

\noindent\textbf{Funding.}
\avela{} faces funding constraints to pay student instructors when scaling its programming to deepen community impact. Compensating student's work is critical for recruiting Black and Latine youth who disproportionately come from low-income backgrounds and need financial support to balance teaching with academic demands and jobs. This requires the nonprofit to substantially supplement the student group's small university grants to maintain consistent programming and recruiting in both college and secondary school settings.

Availability of funding for curriculum materials as well as both virtual and physical tools is critical to providing engaging hands-on lessons on emerging technologies. Hands-on activities also require bringing college mentors and materials like laptops to schools and community centers. Engaging with low-income undergraduates who may not have cars requires \avela{} to have funds available to cover transportation costs and/or pay for virtual tools that can enable secondary school students to participate remotely.



While \avela{} strives to pragmatically deliver STEM access using free resources, the lack of funding for dedicated staff, materials, transportation and digital infrastructure restricts growth potential. Access to grants, sponsors and donors who recognize \avela{}'s impact could enable the organization to robustly serve more youth and communities in a sustainable and reliable manner.

\subsection{Recommendations for Stakeholders}
The key insight is that \avela{}' model effectively addresses the digital divide by facilitating STEM engagements grounded in community priorities and leveraging academic resources. This blueprint empowers underserved Black and Latine university students to build STEM access programs that also support research.

\noindent\textbf{\avela{} Principle 1: Near-peer mentorship and Interactive \& Experiential Learning.}
Recruit Black and Latine students interested in STEM to serve as near-peer mentors and instructors for local youth programs. To do this effectively, students will need to be trained on how to develop hands-on lessons where youth collaboratively build projects, solve problems, and create technologies. To avoid unengaging lectures, \avela{} incorporates research-based activities, simulations, games, and culturally relevant topics. These include highlighting STEM leaders that represent the student population being taught, incorporating STEM history facts, and showcasing opportunities to support their communities and build generational wealth through STEM careers.

\noindent\textbf{\avela{} Principle 2: Mentor embodied community representation and culturally responsive lesson plans.}
Hire mentors who ethnically, culturally, and linguistically represent the program participants. Shared identity builds trust and motivation, but mentors also need to be trained on culturally responsive teaching practices that leverage community contexts, assets, and issues. Co-designing programming with community partners to align with their cultural values and address local needs also helps survey youth and families on interests and obstacles to inform lesson plans that resonate with lived experiences.

\noindent\textbf{\avela{} Principle 3: Compensated Student-led Community Engagement Leveraging Academic Research.}
Support student leaders in directing programming logistics, partnerships, curricula, and outreach. Center student visions by connecting them with experienced mentors, faculty, and research labs pursuing STEM for social change. Helping mentors develop strong university connections so that they can access resources like grants, meeting space, technology, and research opportunities is also vital. By emphasizing reciprocal learning, student mentors will be able to gain technical experience while local partners will be able to help address community needs.

\subsection{Limitations}

We note our sample size is limited due to the size of \avela{}. Additionally, as \avela{} founders, members, and supporters, our research team had insider likeness biases and relationships with interview subjects that may shape our interpretations of the data. The interview participants are those who responded to an email sent to a listserv within \avela{} over the summer, creating selection bias toward more engaged youth. Broadening recruitment in future studies could diversify perspectives. Similarly, more students completed the pre-class survey than the post-class survey, resulting in lower data validity in the post-survey responses. Additional incentives could improve data and confidence in the conclusions from our quantitative analysis. Lastly, this paper provides an in-depth, contextualized look at how equity concerns emerged and were addressed within one specific program, between the summer of 2019 and the summer of 2023, for a particular population. Findings from this study should be put into this context and not over-generalized.

\section{Conclusion}
This study provided an in-depth examination of the persisting digital divide faced by Black and Latine students from urban communities in accessing STEM education and careers, and proposes guiding principles for addressing these issues. Through a mixed-methods approach including interviews, surveys, and autoethnography, we analyzed the sociotechnical access gaps these students encounter and evaluated the impact of the student-led STEM engagement model implemented by the \avela{} nonprofit. 

Our findings confirm that while physical access barriers have decreased, systemic obstacles around belonging, identity, and support systems still persist. We also show that near-peer mentorship, interactive and experiential learning, mentor embodied community representation, culturally responsive lesson planning, and student-led community engagement leveraging academic research and mentorship can combine into a many-methods model to improve perceptions of and interest in pursuing STEM for Black and Latine students. Ultimately, true progress requires acknowledging past injustices, centering student voices, and valuing community knowledge. We highlight one program design that has shown universities how they can leverage their resources to empower youth, transform STEM culture, and create more diverse and equitable pathways into technical fields. 

\section{Acknowledgments}
This work was supported in part by generous donations to the \avela{} nonprofit, and by an award from Amazon Future Engineers (AFE). We also thank Ngozi Ezeokeke for supporting this work since its inception.

\bibliographystyle{ACM-Reference-Format}
\bibliography{references}


\begin{thebibliography}{65}


\ifx \showCODEN    \undefined \def \showCODEN     #1{\unskip}     \fi
\ifx \showDOI      \undefined \def \showDOI       #1{#1}\fi
\ifx \showISBNx    \undefined \def \showISBNx     #1{\unskip}     \fi
\ifx \showISBNxiii \undefined \def \showISBNxiii  #1{\unskip}     \fi
\ifx \showISSN     \undefined \def \showISSN      #1{\unskip}     \fi
\ifx \showLCCN     \undefined \def \showLCCN      #1{\unskip}     \fi
\ifx \shownote     \undefined \def \shownote      #1{#1}          \fi
\ifx \showarticletitle \undefined \def \showarticletitle #1{#1}   \fi
\ifx \showURL      \undefined \def \showURL       {\relax}        \fi
\providecommand\bibfield[2]{#2}
\providecommand\bibinfo[2]{#2}
\providecommand\natexlab[1]{#1}
\providecommand\showeprint[2][]{arXiv:#2}

\bibitem[Alias and Luaran(2016)]%
        {alias2016student}
\bibfield{author}{\bibinfo{person}{Nor~Aziah Alias} {and}
  \bibinfo{person}{Johan~Eddy Luaran}.} \bibinfo{year}{2016}\natexlab{}.
\newblock \bibinfo{booktitle}{\emph{Student-Driven Learning Strategies for the
  21st Century Classroom}}.
\newblock \bibinfo{publisher}{IGI Global}.
\newblock
\showISBNx{9781522516903}
\showLCCN{2016041781}
\urldef\tempurl%
\url{https://books.google.com/books?id=J7WADQAAQBAJ}
\showURL{%
\tempurl}


\bibitem[Amunga(2021)]%
        {levg-tech-stem}
\bibfield{author}{\bibinfo{person}{Jane Amunga}.}
  \bibinfo{year}{2021}\natexlab{}.
\newblock \showarticletitle{Leveraging Technology to Enhance Stem Education
  amidst the Covid-19 Pandemic: An Overview of Pertinent Concerns Education}.
\newblock \bibinfo{journal}{\emph{Technium Social Sciences Journal}}
  \bibinfo{volume}{18} (\bibinfo{year}{2021}).
\newblock
\urldef\tempurl%
\url{https://heinonline.org/HOL/P?h=hein.journals/techssj18&i=40}
\showURL{%
\tempurl}


\bibitem[Anderson et~al\mbox{.}(2019)]%
        {anderson2019benefits}
\bibfield{author}{\bibinfo{person}{Margery~K Anderson},
  \bibinfo{person}{R~Jerome Anderson}, \bibinfo{person}{Laura~S Tenenbaum},
  \bibinfo{person}{Emily~D Kuehn}, \bibinfo{person}{Holly~KM Brown},
  \bibinfo{person}{Swati~B Ramadorai}, {and} \bibinfo{person}{Debra~L
  Yourick}.} \bibinfo{year}{2019}\natexlab{}.
\newblock \showarticletitle{The Benefits of a near-peer mentoring experience on
  STEM persistence in education and careers: A 2004-2015 study}.
\newblock \bibinfo{journal}{\emph{Journal of STEM Outreach}}
  \bibinfo{volume}{2}, \bibinfo{number}{1} (\bibinfo{year}{2019}),
  \bibinfo{pages}{1--11}.
\newblock


\bibitem[Anderson-Butcher and Conroy(2002)]%
        {belonging}
\bibfield{author}{\bibinfo{person}{Dawn Anderson-Butcher} {and}
  \bibinfo{person}{David~E. Conroy}.} \bibinfo{year}{2002}\natexlab{}.
\newblock \showarticletitle{Factorial and Criterion Validity of Scores of a
  Measure of Belonging in Youth Development Programs}.
\newblock \bibinfo{journal}{\emph{Educational and Psychological Measurement}}
  \bibinfo{volume}{62}, \bibinfo{number}{5} (\bibinfo{year}{2002}),
  \bibinfo{pages}{857--876}.
\newblock
\urldef\tempurl%
\url{https://doi.org/10.1177/001316402236882}
\showDOI{\tempurl}


\bibitem[Arroyos et~al\mbox{.}(2022)]%
        {tale}
\bibfield{author}{\bibinfo{person}{Vicente Arroyos}, \bibinfo{person}{Maria~LK
  Viitaniemi}, \bibinfo{person}{Nicholas Keehn}, \bibinfo{person}{Vaidehi
  Oruganti}, \bibinfo{person}{Winston Saunders}, \bibinfo{person}{Karin
  Strauss}, \bibinfo{person}{Vikram Iyer}, {and} \bibinfo{person}{Bichlien~H
  Nguyen}.} \bibinfo{year}{2022}\natexlab{}.
\newblock \showarticletitle{A Tale of Two Mice: Sustainable Electronics Design
  and Prototyping}. In \bibinfo{booktitle}{\emph{CHI Conference on Human
  Factors in Computing Systems Extended Abstracts}}. \bibinfo{pages}{1--10}.
\newblock


\bibitem[Bergerson et~al\mbox{.}(2014)]%
        {bergerson2014outreach}
\bibfield{author}{\bibinfo{person}{Amy~Aldous Bergerson},
  \bibinfo{person}{Bryan~K Hotchkins}, {and} \bibinfo{person}{Cynthia Furse}.}
  \bibinfo{year}{2014}\natexlab{}.
\newblock \showarticletitle{Outreach and identity development: New perspectives
  on college student persistence}.
\newblock \bibinfo{journal}{\emph{Journal of College Student Retention:
  Research, Theory \& Practice}} \bibinfo{volume}{16}, \bibinfo{number}{2}
  (\bibinfo{year}{2014}), \bibinfo{pages}{165--185}.
\newblock


\bibitem[Bourdieu et~al\mbox{.}(1986)]%
        {bourdieu1986handbook}
\bibfield{author}{\bibinfo{person}{Pierre Bourdieu}, \bibinfo{person}{John~G
  Richardson}, {et~al\mbox{.}}} \bibinfo{year}{1986}\natexlab{}.
\newblock \showarticletitle{Handbook of Theory and Research for the Sociology
  of Education}.
\newblock \bibinfo{journal}{\emph{The forms of capital}}  \bibinfo{volume}{241}
  (\bibinfo{year}{1986}), \bibinfo{pages}{258}.
\newblock


\bibitem[Bringle and Hatcher(1996)]%
        {service-learning}
\bibfield{author}{\bibinfo{person}{Robert~G. Bringle} {and}
  \bibinfo{person}{Julie~A. Hatcher}.} \bibinfo{year}{1996}\natexlab{}.
\newblock \showarticletitle{Implementing Service Learning in Higher Education}.
\newblock \bibinfo{journal}{\emph{The Journal of Higher Education}}
  \bibinfo{volume}{67}, \bibinfo{number}{2} (\bibinfo{year}{1996}),
  \bibinfo{pages}{221--239}.
\newblock
\urldef\tempurl%
\url{https://doi.org/10.1080/00221546.1996.11780257}
\showDOI{\tempurl}


\bibitem[Caselman et~al\mbox{.}({[n.\,d.]})]%
        {is-hs}
\bibfield{author}{\bibinfo{person}{Tonia~D. Caselman},
  \bibinfo{person}{Patricia~A. Self}, {and} \bibinfo{person}{Angela~L. Self}.}
  \bibinfo{year}{[n.\,d.]}\natexlab{}.
\newblock \showarticletitle{Adolescent attributes contributing to the imposter
  phenomenon☆}.
\newblock \bibinfo{journal}{\emph{Journal of Adolescence}}
  \bibinfo{volume}{29}, \bibinfo{number}{3} (\bibinfo{year}{[n.\,d.]}),
  \bibinfo{pages}{395--405}.
\newblock
\urldef\tempurl%
\url{https://doi.org/10.1016/j.adolescence.2005.07.003}
\showDOI{\tempurl}


\bibitem[Chen et~al\mbox{.}(2023)]%
        {Chen2023}
\bibfield{author}{\bibinfo{person}{Chen Chen}, \bibinfo{person}{Jonathan
  Rothwell}, {and} \bibinfo{person}{pedrito~maynard zhang}.}
  \bibinfo{year}{2023}\natexlab{}.
\newblock \showarticletitle{In-school and/or out-of-school computer science
  learning influence on CS career interests, mediated by having role-models}.
  In \bibinfo{booktitle}{\emph{AERA 2023 Workshop on Multimodal Literacy in OST
  Programs: Family and Community Ties}}.
\newblock
\urldef\tempurl%
\url{https://www.amazon.science/publications/in-school-and-or-out-of-school-computer-science-learning-influence-on-cs-career-interests-mediated-by-having-role-models}
\showURL{%
\tempurl}


\bibitem[Chordia(2022)]%
        {tj-chi}
\bibfield{author}{\bibinfo{person}{Ishita Chordia}.}
  \bibinfo{year}{2022}\natexlab{}.
\newblock \showarticletitle{Leveraging Transformative Justice in Organizing
  Collective Action Towards Community Safety} \emph{(\bibinfo{series}{CHI EA
  '22})}. \bibinfo{publisher}{Association for Computing Machinery},
  \bibinfo{address}{New York, NY, USA}, Article \bibinfo{articleno}{50},
  \bibinfo{numpages}{4}~pages.
\newblock
\showISBNx{9781450391566}
\urldef\tempurl%
\url{https://doi.org/10.1145/3491101.3503820}
\showDOI{\tempurl}


\bibitem[Commission(2022)]%
        {eeoc2022}
\bibfield{author}{\bibinfo{person}{U.S. Equal Employment~Opportunity
  Commission}.} \bibinfo{year}{2022}\natexlab{}.
\newblock \bibinfo{title}{EEOC Fiscal Year 2022 Agency Financial Report}.
\newblock
\newblock
\urldef\tempurl%
\url{https://www.eeoc.gov/eeoc-fiscal-year-2022-agency-financial-report}
\showURL{%
\tempurl}


\bibitem[Costanza-Chock(2018)]%
        {costanza2018design}
\bibfield{author}{\bibinfo{person}{Sasha Costanza-Chock}.}
  \bibinfo{year}{2018}\natexlab{}.
\newblock \showarticletitle{Design justice: Towards an intersectional feminist
  framework for design theory and practice}.
\newblock \bibinfo{journal}{\emph{Proceedings of the Design Research Society}}
  (\bibinfo{year}{2018}).
\newblock


\bibitem[Cunningham(2023)]%
        {cunningham2023collaboratively}
\bibfield{author}{\bibinfo{person}{Jay~L Cunningham}.}
  \bibinfo{year}{2023}\natexlab{}.
\newblock \showarticletitle{Collaboratively Mitigating Racial Disparities in
  Automated Speech Recognition and Language Technologies with African American
  English Speakers: Community-Collaborative and Equity-Centered Approaches
  Toward Designing Inclusive Natural Language Systems}. In
  \bibinfo{booktitle}{\emph{Extended Abstracts of the 2023 CHI Conference on
  Human Factors in Computing Systems}}. \bibinfo{pages}{1--5}.
\newblock


\bibitem[Cunningham et~al\mbox{.}(2022)]%
        {cunningham2022cost}
\bibfield{author}{\bibinfo{person}{Jay~L Cunningham}, \bibinfo{person}{Sydney~T
  Nguyen}, \bibinfo{person}{Julie~A Kientz}, {and} \bibinfo{person}{Daniela
  Rosner}.} \bibinfo{year}{2022}\natexlab{}.
\newblock \showarticletitle{The Cost of Culture: An Analysis of Cash App and
  the Financial Inclusion of Black American Communities}. In
  \bibinfo{booktitle}{\emph{Designing Interactive Systems Conference}}.
  \bibinfo{pages}{612--628}.
\newblock


\bibitem[DiSalvo and DesPortes(2017)]%
        {disalvo2017participatory}
\bibfield{author}{\bibinfo{person}{Betsy DiSalvo} {and} \bibinfo{person}{Kayla
  DesPortes}.} \bibinfo{year}{2017}\natexlab{}.
\newblock \bibinfo{booktitle}{\emph{Participatory Design for Value-Driven
  Learning} (\bibinfo{edition}{1st} ed.)}.
\newblock \bibinfo{publisher}{Routledge}, \bibinfo{pages}{14}.
\newblock
\showISBNx{9781315630830}


\bibitem[DiSalvo et~al\mbox{.}(2014)]%
        {disalvo2014saving}
\bibfield{author}{\bibinfo{person}{Betsy DiSalvo}, \bibinfo{person}{Mark
  Guzdial}, \bibinfo{person}{Amy Bruckman}, {and} \bibinfo{person}{Tom
  McKlin}.} \bibinfo{year}{2014}\natexlab{}.
\newblock \showarticletitle{Saving face while geeking out: Video game testing
  as a justification for learning computer science}.
\newblock \bibinfo{journal}{\emph{Journal of the Learning Sciences}}
  \bibinfo{volume}{23}, \bibinfo{number}{3} (\bibinfo{year}{2014}),
  \bibinfo{pages}{272--315}.
\newblock


\bibitem[Erete et~al\mbox{.}(2021)]%
        {transformative-justice}
\bibfield{author}{\bibinfo{person}{Sheena Erete}, \bibinfo{person}{Karla
  Thomas}, \bibinfo{person}{Denise Nacu}, \bibinfo{person}{Jessa Dickinson},
  \bibinfo{person}{Naomi Thompson}, {and} \bibinfo{person}{Nichole Pinkard}.}
  \bibinfo{year}{2021}\natexlab{}.
\newblock \showarticletitle{Applying a Transformative Justice Approach to
  Encourage the Participation of Black and Latina Girls in Computing}.
\newblock \bibinfo{journal}{\emph{ACM Trans. Comput. Educ.}}
  \bibinfo{volume}{21}, \bibinfo{number}{4}, Article \bibinfo{articleno}{27}
  (\bibinfo{date}{oct} \bibinfo{year}{2021}), \bibinfo{numpages}{24}~pages.
\newblock
\urldef\tempurl%
\url{https://doi.org/10.1145/3451345}
\showDOI{\tempurl}


\bibitem[{Federal Student Aid}(2023)]%
        {pell}
\bibfield{author}{\bibinfo{person}{{Federal Student Aid}}.}
  \bibinfo{year}{2023}\natexlab{}.
\newblock \bibinfo{title}{Federal Pell Grants}.
\newblock
\newblock
\urldef\tempurl%
\url{https://studentaid.gov/understand-aid/types/grants/pell}
\showURL{%
\tempurl}


\bibitem[Foundation(2023)]%
        {nsf-access}
\bibfield{author}{\bibinfo{person}{National~Science Foundation}.}
  \bibinfo{year}{2023}\natexlab{}.
\newblock \bibinfo{title}{NSF Public Access Plan 2.0}.
\newblock
\newblock
\urldef\tempurl%
\url{https://www.nsf.gov/pubs/2023/nsf23104/nsf23104.pdf}
\showURL{%
\tempurl}


\bibitem[Freeman and Jurow(2018)]%
        {disruptive-teaching}
\bibfield{author}{\bibinfo{person}{Quinton Freeman} {and} \bibinfo{person}{A.
  Jurow}.} \bibinfo{year}{2018}\natexlab{}.
\newblock \bibinfo{booktitle}{\emph{Becoming a More Disruptive Teacher by
  Engaging in Side-by-Side Learning with Children Rather than Avoiding
  Discomfort}}.
\newblock \bibinfo{pages}{35--52}.
\newblock


\bibitem[Gay(2002)]%
        {culturally-responsive}
\bibfield{author}{\bibinfo{person}{Geneva Gay}.}
  \bibinfo{year}{2002}\natexlab{}.
\newblock \showarticletitle{Preparing for Culturally Responsive Teaching}.
\newblock \bibinfo{journal}{\emph{Journal of Teacher Education}}
  \bibinfo{volume}{53}, \bibinfo{number}{2} (\bibinfo{year}{2002}),
  \bibinfo{pages}{106--116}.
\newblock
\urldef\tempurl%
\url{https://doi.org/10.1177/0022487102053002003}
\showDOI{\tempurl}


\bibitem[Grabsch et~al\mbox{.}(2023)]%
        {grabsch2023using}
\bibfield{author}{\bibinfo{person}{Dustin~K Grabsch}, \bibinfo{person}{Lori~L
  Moore}, \bibinfo{person}{Meredith Levesque}, {and} \bibinfo{person}{Taelor
  Robinson}.} \bibinfo{year}{2023}\natexlab{}.
\newblock \showarticletitle{Using Community Cultural Wealth: An Asset-Based
  Approach to Persistence of On-Campus Black and Latinx Collegians}.
\newblock \bibinfo{journal}{\emph{Journal of College Student Retention:
  Research, Theory \& Practice}} (\bibinfo{year}{2023}),
  \bibinfo{pages}{15210251231192884}.
\newblock


\bibitem[Graham et~al\mbox{.}(2013)]%
        {graham2013}
\bibfield{author}{\bibinfo{person}{Mark~J. Graham}, \bibinfo{person}{Jennifer
  Frederick}, \bibinfo{person}{Angela Byars-Winston},
  \bibinfo{person}{Anne-Barrie Hunter}, {and} \bibinfo{person}{Jo Handelsman}.}
  \bibinfo{year}{2013}\natexlab{}.
\newblock \showarticletitle{Increasing Persistence of College Students in
  {STEM}}.
\newblock \bibinfo{journal}{\emph{Science}}  \bibinfo{volume}{341}
  (\bibinfo{year}{2013}), \bibinfo{pages}{1455--1456}.
\newblock
\urldef\tempurl%
\url{https://doi.org/10.1126/science.1240487}
\showDOI{\tempurl}


\bibitem[Gutiérrez and Jurow(2016)]%
        {equity-design}
\bibfield{author}{\bibinfo{person}{Kris~D. Gutiérrez} {and}
  \bibinfo{person}{A.~Susan Jurow}.} \bibinfo{year}{2016}\natexlab{}.
\newblock \showarticletitle{Social Design Experiments: Toward Equity by
  Design}.
\newblock \bibinfo{journal}{\emph{Journal of the Learning Sciences}}
  \bibinfo{volume}{25}, \bibinfo{number}{4} (\bibinfo{year}{2016}),
  \bibinfo{pages}{565--598}.
\newblock
\urldef\tempurl%
\url{https://doi.org/10.1080/10508406.2016.1204548}
\showDOI{\tempurl}


\bibitem[Harper(2010)]%
        {harper2010anti}
\bibfield{author}{\bibinfo{person}{Shaun~R Harper}.}
  \bibinfo{year}{2010}\natexlab{}.
\newblock \showarticletitle{An anti-deficit achievement framework for research
  on students of color in STEM}.
\newblock \bibinfo{journal}{\emph{New directions for institutional research}}
  \bibinfo{volume}{2010}, \bibinfo{number}{148} (\bibinfo{year}{2010}),
  \bibinfo{pages}{63--74}.
\newblock


\bibitem[Harrington et~al\mbox{.}(2019)]%
        {harrington2019deconstructing}
\bibfield{author}{\bibinfo{person}{Christina Harrington},
  \bibinfo{person}{Sheena Erete}, {and} \bibinfo{person}{Anne~Marie Piper}.}
  \bibinfo{year}{2019}\natexlab{}.
\newblock \showarticletitle{Deconstructing community-based collaborative
  design: Towards more equitable participatory design engagements}.
\newblock \bibinfo{journal}{\emph{Proceedings of the ACM on Human-Computer
  Interaction}} \bibinfo{volume}{3}, \bibinfo{number}{CSCW}
  (\bibinfo{year}{2019}), \bibinfo{pages}{1--25}.
\newblock


\bibitem[Hotchkins et~al\mbox{.}(2019)]%
        {hotchkins2019first}
\bibfield{author}{\bibinfo{person}{Bryan Hotchkins}, \bibinfo{person}{Nedra
  Hotchkins}, \bibinfo{person}{Bianca Bellot}, {and} \bibinfo{person}{Laurence
  Parker}.} \bibinfo{year}{2019}\natexlab{}.
\newblock \showarticletitle{First year college student success for Black and
  other students of color: A Village initiative at the University of Utah}.
\newblock \bibinfo{journal}{\emph{Journal of Curriculum, Teaching, Learning and
  Leadership in Education}} \bibinfo{volume}{4}, \bibinfo{number}{1}
  (\bibinfo{year}{2019}), \bibinfo{pages}{28}.
\newblock


\bibitem[Kang and Nation(2023)]%
        {co-design-action}
\bibfield{author}{\bibinfo{person}{Hosun Kang} {and}
  \bibinfo{person}{Jasmine~McBeath Nation}.} \bibinfo{year}{2023}\natexlab{}.
\newblock \showarticletitle{Transforming Science Learning Framework:
  Translating an Equity Commitment into Action through Co-Design}.
\newblock \bibinfo{journal}{\emph{Journal of Science Teacher Education}}
  \bibinfo{volume}{34}, \bibinfo{number}{6} (\bibinfo{year}{2023}),
  \bibinfo{pages}{667--687}.
\newblock
\urldef\tempurl%
\url{https://doi.org/10.1080/1046560X.2022.2132633}
\showDOI{\tempurl}


\bibitem[Krueger and Whitmore(2001)]%
        {krueger2001effect}
\bibfield{author}{\bibinfo{person}{Alan~B Krueger} {and}
  \bibinfo{person}{Diane~M Whitmore}.} \bibinfo{year}{2001}\natexlab{}.
\newblock \showarticletitle{The effect of attending a small class in the early
  grades on college-test taking and middle school test results: Evidence from
  Project STAR}.
\newblock \bibinfo{journal}{\emph{The Economic Journal}} \bibinfo{volume}{111},
  \bibinfo{number}{468} (\bibinfo{year}{2001}), \bibinfo{pages}{1--28}.
\newblock


\bibitem[Lachney et~al\mbox{.}(2021)]%
        {aa-cosmetology}
\bibfield{author}{\bibinfo{person}{Michael Lachney}, \bibinfo{person}{William
  Babbitt}, \bibinfo{person}{Audrey Bennett}, {and} \bibinfo{person}{Ron
  Eglash}.} \bibinfo{year}{2021}\natexlab{}.
\newblock \showarticletitle{Generative computing: African-American cosmetology
  as a link between computing education and community wealth}.
\newblock \bibinfo{journal}{\emph{Interactive Learning Environments}}
  \bibinfo{volume}{29}, \bibinfo{number}{7} (\bibinfo{year}{2021}),
  \bibinfo{pages}{1115--1135}.
\newblock
\urldef\tempurl%
\url{https://doi.org/10.1080/10494820.2019.1636087}
\showDOI{\tempurl}


\bibitem[Ladson-Billings(1995)]%
        {cr-pedagogy}
\bibfield{author}{\bibinfo{person}{Gloria Ladson-Billings}.}
  \bibinfo{year}{1995}\natexlab{}.
\newblock \showarticletitle{Toward a Theory of Culturally Relevant Pedagogy}.
\newblock \bibinfo{journal}{\emph{American Educational Research Journal}}
  \bibinfo{volume}{32}, \bibinfo{number}{3} (\bibinfo{year}{1995}),
  \bibinfo{pages}{465--491}.
\newblock
\showISSN{00028312, 19351011}
\urldef\tempurl%
\url{http://www.jstor.org/stable/1163320}
\showURL{%
\tempurl}


\bibitem[Luederitz et~al\mbox{.}(2016)]%
        {LUEDERITZ2016229}
\bibfield{author}{\bibinfo{person}{Christopher Luederitz},
  \bibinfo{person}{Moritz Meyer}, \bibinfo{person}{David~J. Abson},
  \bibinfo{person}{Fabienne Gralla}, \bibinfo{person}{Daniel~J. Lang},
  \bibinfo{person}{Anna-Lena Rau}, {and} \bibinfo{person}{Henrik {von
  Wehrden}}.} \bibinfo{year}{2016}\natexlab{}.
\newblock \showarticletitle{Systematic student-driven literature reviews in
  sustainability science – an effective way to merge research and teaching}.
\newblock \bibinfo{journal}{\emph{Journal of Cleaner Production}}
  \bibinfo{volume}{119} (\bibinfo{year}{2016}), \bibinfo{pages}{229--235}.
\newblock
\showISSN{0959-6526}
\urldef\tempurl%
\url{https://doi.org/10.1016/j.jclepro.2016.02.005}
\showDOI{\tempurl}


\bibitem[Lusk et~al\mbox{.}(2020)]%
        {ccw-chi}
\bibfield{author}{\bibinfo{person}{Gena Lusk}, \bibinfo{person}{Trenia Walker},
  {and} \bibinfo{person}{Holly Ferraro}.} \bibinfo{year}{2020}\natexlab{}.
\newblock \emph{\bibinfo{title}{Community Cultural Wealth in the Community
  College: A Systematic Review of Latinx Student Engagement}}.
\newblock \bibinfo{thesistype}{Ph.\,D. Dissertation}.
\newblock Advisor(s) M, Taylor, Colette.
\newblock
\showISBNx{9798662384484}


\bibitem[Margolis(2017)]%
        {margolis2017stuck}
\bibfield{author}{\bibinfo{person}{Jane Margolis}.}
  \bibinfo{year}{2017}\natexlab{}.
\newblock \bibinfo{booktitle}{\emph{Stuck in the shallow end, updated edition:
  Education, race, and computing}}.
\newblock \bibinfo{publisher}{MIT press}.
\newblock


\bibitem[Mawasi et~al\mbox{.}(2023)]%
        {co-design-process}
\bibfield{author}{\bibinfo{person}{Areej Mawasi}, \bibinfo{person}{William
  Penuel}, \bibinfo{person}{Arturo Cortez}, {and} \bibinfo{person}{Ashieda
  McKoy}.} \bibinfo{year}{2023}\natexlab{}.
\newblock \showarticletitle{“They were learning from us as we were learning
  from them”: perceived experiences in co-design process}.
\newblock \bibinfo{journal}{\emph{Mind, Culture, and Activity}}
  \bibinfo{volume}{0}, \bibinfo{number}{0} (\bibinfo{year}{2023}),
  \bibinfo{pages}{1--18}.
\newblock
\urldef\tempurl%
\url{https://doi.org/10.1080/10749039.2023.2246039}
\showDOI{\tempurl}


\bibitem[McGee and Bentley(2017)]%
        {bw-stem}
\bibfield{author}{\bibinfo{person}{Ebony McGee} {and} \bibinfo{person}{Lydia
  Bentley}.} \bibinfo{year}{2017}\natexlab{}.
\newblock \showarticletitle{The Troubled Success of Black Women in STEM}.
\newblock \bibinfo{journal}{\emph{Cognition and Instruction}}
  (\bibinfo{date}{08} \bibinfo{year}{2017}), \bibinfo{pages}{1--25}.
\newblock
\urldef\tempurl%
\url{https://doi.org/10.1080/07370008.2017.1355211}
\showDOI{\tempurl}


\bibitem[of~Black~Engineers(2021)]%
        {nsbe-demographics}
\bibfield{author}{\bibinfo{person}{National~Society of Black~Engineers}.}
  \bibinfo{year}{2021}\natexlab{}.
\newblock \bibinfo{title}{Advancing the Black STEM Experience}.
\newblock
\newblock
\urldef\tempurl%
\url{https://www.nsbe.org/getmedia/620397c3-118e-4da6-a0af-ce401bb527a2/NSBE-2021%e2%80%932022-Annual-Report.pdf}
\showURL{%
\tempurl}


\bibitem[of~Hispanic Professional~Engineers(2022)]%
        {shpe-demographics}
\bibfield{author}{\bibinfo{person}{Society of Hispanic
  Professional~Engineers}.} \bibinfo{year}{2022}\natexlab{}.
\newblock \bibinfo{title}{2022 SHPE National Convention: First Look Exhibitor
  Opportunities}.
\newblock
\newblock
\urldef\tempurl%
\url{https://shpe.org/wp-content/uploads/2023/01/SHPE2022-23-demographics.pdf}
\showURL{%
\tempurl}


\bibitem[of~Planning and Budgeting(2023)]%
        {pell_urm_all_UW}
\bibfield{author}{\bibinfo{person}{UW~Office of Planning} {and}
  \bibinfo{person}{Budgeting}.} \bibinfo{year}{2023}\natexlab{}.
\newblock \bibinfo{title}{Undergraduate Student Persistance and Graduation by
  Year of Study occupations: 2019–29}.
\newblock
\newblock
\urldef\tempurl%
\url{https://www.washington.edu/opb/uw-data/uw-profiles-information/}
\showURL{%
\tempurl}


\bibitem[Ovink and Veazey(2011)]%
        {ovink2011more}
\bibfield{author}{\bibinfo{person}{Sarah~M Ovink} {and}
  \bibinfo{person}{Brian~D Veazey}.} \bibinfo{year}{2011}\natexlab{}.
\newblock \showarticletitle{More than “getting us through:” A case study in
  cultural capital enrichment of underrepresented minority undergraduates}.
\newblock \bibinfo{journal}{\emph{Research in higher education}}
  \bibinfo{volume}{52} (\bibinfo{year}{2011}), \bibinfo{pages}{370--394}.
\newblock


\bibitem[Parkman(2016)]%
        {imposter-syndrome}
\bibfield{author}{\bibinfo{person}{Anna Parkman}.}
  \bibinfo{year}{2016}\natexlab{}.
\newblock \showarticletitle{The Imposter Phenomenon in Higher Education:
  Incidence and Impact}.
\newblock \bibinfo{journal}{\emph{Journal of Higher Education Theory and
  Practice}}  \bibinfo{volume}{16} (\bibinfo{date}{01} \bibinfo{year}{2016}),
  \bibinfo{pages}{51--60}.
\newblock


\bibitem[Parnell et~al\mbox{.}(2023)]%
        {parnell2023engineering}
\bibfield{author}{\bibinfo{person}{Dennis~R Parnell}, \bibinfo{person}{Jabari
  Wilson}, \bibinfo{person}{Karen~Theodora Hicklin}, {and}
  \bibinfo{person}{Jeremy A~Magruder Waisome}.}
  \bibinfo{year}{2023}\natexlab{}.
\newblock \showarticletitle{Engineering While Black: Exploring the Experiences
  of Black University of Florida Undergraduate Engineering Students Using
  Photovoice}. In \bibinfo{booktitle}{\emph{2023 ASEE Annual Conference \&
  Exposition}}.
\newblock


\bibitem[Patterson Silver~Wolf et~al\mbox{.}(2021)]%
        {patterson2021you}
\bibfield{author}{\bibinfo{person}{David~A Patterson Silver~Wolf},
  \bibinfo{person}{Franklyn Taylor}, \bibinfo{person}{Eugene Maguin}, {and}
  \bibinfo{person}{Autumn Asher~BlackDeer}.} \bibinfo{year}{2021}\natexlab{}.
\newblock \showarticletitle{You are college material—You belong: An
  underrepresented minority student retention intervention without deception}.
\newblock \bibinfo{journal}{\emph{Journal of College Student Retention:
  Research, Theory \& Practice}} \bibinfo{volume}{23}, \bibinfo{number}{3}
  (\bibinfo{year}{2021}), \bibinfo{pages}{507--522}.
\newblock


\bibitem[Pipa~Stevens(2020)]%
        {digital_racial_gap}
\bibfield{author}{\bibinfo{person}{CNBC Pipa~Stevens}.}
  \bibinfo{year}{2020}\natexlab{}.
\newblock \bibinfo{title}{Digital racial gap could ‘render the country’s
  minorities into an unemployment abyss,’ says Deutsche Bank}.
\newblock
\newblock
\urldef\tempurl%
\url{https://www.cnbc.com/2020/09/02/digital-racial-gap-could-render-the-countrys-minorities-into-an-unemployment-abyss-says-deutsche-bank.html}
\showURL{%
\tempurl}


\bibitem[Pluth et~al\mbox{.}(2015)]%
        {pluth2015collaboration}
\bibfield{author}{\bibinfo{person}{Michael~D Pluth}, \bibinfo{person}{Shannon~W
  Boettcher}, \bibinfo{person}{George~V Nazin}, \bibinfo{person}{Ann~L
  Greenaway}, {and} \bibinfo{person}{Matthew~D Hartle}.}
  \bibinfo{year}{2015}\natexlab{}.
\newblock \showarticletitle{Collaboration and near-peer mentoring as a platform
  for sustainable science education outreach}.
\newblock \bibinfo{journal}{\emph{Journal of Chemical Education}}
  \bibinfo{volume}{92}, \bibinfo{number}{4} (\bibinfo{year}{2015}),
  \bibinfo{pages}{625--630}.
\newblock


\bibitem[Rashid and Yadav(2020)]%
        {covid-education2}
\bibfield{author}{\bibinfo{person}{Shazia Rashid} {and}
  \bibinfo{person}{Sunishtha~Singh Yadav}.} \bibinfo{year}{2020}\natexlab{}.
\newblock \showarticletitle{Impact of Covid-19 Pandemic on Higher Education and
  Research}.
\newblock \bibinfo{journal}{\emph{Indian Journal of Human Development}}
  \bibinfo{volume}{14}, \bibinfo{number}{2} (\bibinfo{year}{2020}),
  \bibinfo{pages}{340--343}.
\newblock
\urldef\tempurl%
\url{https://doi.org/10.1177/0973703020946700}
\showDOI{\tempurl}


\bibitem[Rose et~al\mbox{.}(2020)]%
        {sl-chi}
\bibfield{author}{\bibinfo{person}{Brittney~D. Rose}, \bibinfo{person}{Eddie
  Hill}, {and} \bibinfo{person}{Jori Beck}.} \bibinfo{year}{2020}\natexlab{}.
\newblock \emph{\bibinfo{title}{Service Learning and the Experiential Learning
  Cycle in Elementary School}}.
\newblock \bibinfo{thesistype}{Ph.\,D. Dissertation}.
\newblock Advisor(s) Angela, Eckhoff,.
\newblock
\showISBNx{9781083217004}


\bibitem[Rosenow et~al\mbox{.}(2020)]%
        {cr-chi}
\bibfield{author}{\bibinfo{person}{Jane Alice~Wambui Rosenow},
  \bibinfo{person}{Steven White}, \bibinfo{person}{Zachary Foste},
  \bibinfo{person}{Arlene Barry}, {and} \bibinfo{person}{Carrie~La Voy}.}
  \bibinfo{year}{2020}\natexlab{}.
\newblock \emph{\bibinfo{title}{Culturally Responsive Teaching in Practice:
  Exploring English Language Arts Classrooms That Serve Ethnically,
  Linguistically, and Culturally Diverse Students}}.
\newblock \bibinfo{thesistype}{Ph.\,D. Dissertation}.
\newblock Advisor(s) Heidi, Hallman,.
\newblock
\showISBNx{9798662417021}


\bibitem[Ross and McGrade(2016)]%
        {nsbe-27280}
\bibfield{author}{\bibinfo{person}{Monique~S. Ross} {and}
  \bibinfo{person}{Susan McGrade}.} \bibinfo{year}{2016}\natexlab{}.
\newblock \showarticletitle{An Exploration into the Impacts of the National
  Society of Black Engineers (NSBE) on Student Persistence}. In
  \bibinfo{booktitle}{\emph{2016 ASEE Annual Conference \& Exposition}}.
  \bibinfo{publisher}{ASEE Conferences}, \bibinfo{address}{New Orleans,
  Louisiana}.
\newblock
\urldef\tempurl%
\url{https://doi.org/10.18260/p.27280}
\showDOI{\tempurl}
\newblock
\shownote{https://peer.asee.org/27280}.


\bibitem[SAM~DEAN(2020)]%
        {kept_out}
\bibfield{author}{\bibinfo{person}{JOHANA~BHUIYAN SAM~DEAN}.}
  \bibinfo{year}{2020}\natexlab{}.
\newblock \bibinfo{title}{Why are Black and Latino people still kept out of the
  tech industry?}
\newblock
\newblock
\urldef\tempurl%
\url{https://www.latimes.com/business/technology/story/2020-06-24/tech-started-publicly-taking-lack-of-diversity-seriously-in-2014-why-has-so-little-changed-for-black-workers}
\showURL{%
\tempurl}


\bibitem[Samuelson and Litzler(2016a)]%
        {ccw}
\bibfield{author}{\bibinfo{person}{Cate~C. Samuelson} {and}
  \bibinfo{person}{Elizabeth Litzler}.} \bibinfo{year}{2016}\natexlab{a}.
\newblock \showarticletitle{Community Cultural Wealth: An Assets-Based Approach
  to Persistence of Engineering Students of Color}.
\newblock \bibinfo{journal}{\emph{Journal of Engineering Education}}
  \bibinfo{volume}{105}, \bibinfo{number}{1} (\bibinfo{year}{2016}),
  \bibinfo{pages}{93--117}.
\newblock
\urldef\tempurl%
\url{https://doi.org/10.1002/jee.20110}
\showDOI{\tempurl}


\bibitem[Samuelson and Litzler(2016b)]%
        {samuelson2016community}
\bibfield{author}{\bibinfo{person}{Cate~C Samuelson} {and}
  \bibinfo{person}{Elizabeth Litzler}.} \bibinfo{year}{2016}\natexlab{b}.
\newblock \showarticletitle{Community cultural wealth: An assets-based approach
  to persistence of engineering students of color}.
\newblock \bibinfo{journal}{\emph{Journal of Engineering Education}}
  \bibinfo{volume}{105}, \bibinfo{number}{1} (\bibinfo{year}{2016}),
  \bibinfo{pages}{93--117}.
\newblock


\bibitem[Sanders and Edward(2021)]%
        {digital-divide}
\bibfield{author}{\bibinfo{person}{Scanlon Sanders, Cynthia~K.} {and}
  \bibinfo{person}{Edward}.} \bibinfo{year}{2021}\natexlab{}.
\newblock \showarticletitle{The Digital Divide Is a Human Rights Issue:
  Advancing Social Inclusion Through Social Work Advocacy}.
\newblock \bibinfo{journal}{\emph{Journal of Human Rights and Social Work}}
  \bibinfo{volume}{6}, \bibinfo{number}{2} (\bibinfo{year}{2021}),
  \bibinfo{pages}{2365--1792}.
\newblock
\urldef\tempurl%
\url{https://doi.org/10.1007/s41134-020-00147-9}
\showDOI{\tempurl}


\bibitem[Singer et~al\mbox{.}(2020)]%
        {singer2020}
\bibfield{author}{\bibinfo{person}{Alison Singer}, \bibinfo{person}{Georgina
  Montgomery}, {and} \bibinfo{person}{Shannon Schmoll}.}
  \bibinfo{year}{2020}\natexlab{}.
\newblock \showarticletitle{How to foster the formation of STEM identity:
  studying diversity in an authentic learning environment}.
\newblock \bibinfo{journal}{\emph{International Journal of STEM Education}}
  \bibinfo{volume}{7}, \bibinfo{number}{1} (\bibinfo{date}{11}
  \bibinfo{year}{2020}).
\newblock
\urldef\tempurl%
\url{https://doi.org/10.1186/s40594-020-00254-z}
\showDOI{\tempurl}


\bibitem[Syed et~al\mbox{.}(2012)]%
        {syed2012individual}
\bibfield{author}{\bibinfo{person}{Moin Syed}, \bibinfo{person}{Barbara~K
  Goza}, \bibinfo{person}{Martin~M Chemers}, {and} \bibinfo{person}{Eileen~L
  Zurbriggen}.} \bibinfo{year}{2012}\natexlab{}.
\newblock \showarticletitle{Individual differences in preferences for
  matched-ethnic mentors among high-achieving ethnically diverse adolescents in
  STEM}.
\newblock \bibinfo{journal}{\emph{Child development}} \bibinfo{volume}{83},
  \bibinfo{number}{3} (\bibinfo{year}{2012}), \bibinfo{pages}{896--910}.
\newblock


\bibitem[Tadesse and Muluye(2020)]%
        {covid-education1}
\bibfield{author}{\bibinfo{person}{S. Tadesse} {and} \bibinfo{person}{W.
  Muluye}.} \bibinfo{year}{2020}\natexlab{}.
\newblock \showarticletitle{The Impact of COVID-19 Pandemic on Education System
  in Developing Countries: A Review}.
\newblock \bibinfo{journal}{\emph{Open Journal of Social Sciences}}
  \bibinfo{volume}{8} (\bibinfo{year}{2020}), \bibinfo{pages}{159--170}.
\newblock
\urldef\tempurl%
\url{https://doi.org/10.4236/jss.2020.810011}
\showDOI{\tempurl}


\bibitem[Trenshaw et~al\mbox{.}(2020)]%
        {role-models}
\bibfield{author}{\bibinfo{person}{Kyle Trenshaw}, \bibinfo{person}{Derek
  Rushton}, \bibinfo{person}{Elif~Eda Miskio{\u{g}}lu}, {and}
  \bibinfo{person}{Philip Asare}.} \bibinfo{year}{2020}\natexlab{}.
\newblock \showarticletitle{Work-in-Progress: Emergent Themes from" High
  Impact" Role Model and Mentor Narratives}. In \bibinfo{booktitle}{\emph{2020
  IEEE Frontiers in Education Conference (FIE)}}. IEEE, \bibinfo{pages}{1--4}.
\newblock


\bibitem[Villalpando and Sol{\'o}rzano(2005)]%
        {villalpando2005role}
\bibfield{author}{\bibinfo{person}{Octavio Villalpando} {and}
  \bibinfo{person}{Daniel~G Sol{\'o}rzano}.} \bibinfo{year}{2005}\natexlab{}.
\newblock \showarticletitle{The role of culture in college preparation
  programs: A review of the research literature}.
\newblock \bibinfo{journal}{\emph{Preparing for college: Nine elements of
  effective outreach}} (\bibinfo{year}{2005}), \bibinfo{pages}{13--28}.
\newblock


\bibitem[Washington and Johnson(2023)]%
        {shanee1}
\bibfield{author}{\bibinfo{person}{Shaneé~A. Washington} {and}
  \bibinfo{person}{Lauri Johnson}.} \bibinfo{year}{2023}\natexlab{}.
\newblock \showarticletitle{Toward culturally sustaining/revitalizing
  Indigenous family-school-community leadership}.
\newblock \bibinfo{journal}{\emph{Frontiers in Education}}  \bibinfo{volume}{8}
  (\bibinfo{year}{2023}).
\newblock
\showISSN{2504-284X}
\urldef\tempurl%
\url{https://doi.org/10.3389/feduc.2023.1192095}
\showDOI{\tempurl}


\bibitem[Wong-Villacres et~al\mbox{.}(2020)]%
        {wong2020needs}
\bibfield{author}{\bibinfo{person}{Marisol Wong-Villacres},
  \bibinfo{person}{Aakash Gautam}, \bibinfo{person}{Wendy Roldan},
  \bibinfo{person}{Lucy Pei}, \bibinfo{person}{Jessa Dickinson},
  \bibinfo{person}{Azra Ismail}, \bibinfo{person}{Betsy DiSalvo},
  \bibinfo{person}{Neha Kumar}, \bibinfo{person}{Tammy Clegg},
  \bibinfo{person}{Sheena Erete}, {et~al\mbox{.}}}
  \bibinfo{year}{2020}\natexlab{}.
\newblock \showarticletitle{From needs to strengths: Operationalizing an
  assets-based design of technology}. In \bibinfo{booktitle}{\emph{Conference
  Companion Publication of the 2020 on Computer Supported Cooperative Work and
  Social Computing}}. \bibinfo{pages}{527--535}.
\newblock


\bibitem[Wright and Wright(1987)]%
        {wright1987role}
\bibfield{author}{\bibinfo{person}{Cheryl~A Wright} {and}
  \bibinfo{person}{Scott~D Wright}.} \bibinfo{year}{1987}\natexlab{}.
\newblock \showarticletitle{The role of mentors in the career development of
  young professionals}.
\newblock \bibinfo{journal}{\emph{Family relations}} (\bibinfo{year}{1987}),
  \bibinfo{pages}{204--208}.
\newblock


\bibitem[Yosso*(2005)]%
        {yosso2005whose}
\bibfield{author}{\bibinfo{person}{Tara~J Yosso*}.}
  \bibinfo{year}{2005}\natexlab{}.
\newblock \showarticletitle{Whose culture has capital? A critical race theory
  discussion of community cultural wealth}.
\newblock \bibinfo{journal}{\emph{Race ethnicity and education}}
  \bibinfo{volume}{8}, \bibinfo{number}{1} (\bibinfo{year}{2005}),
  \bibinfo{pages}{69--91}.
\newblock


\bibitem[Young et~al\mbox{.}(2014)]%
        {co-curricular}
\bibfield{author}{\bibinfo{person}{Glenda Young}, \bibinfo{person}{David~B.
  Knight}, {and} \bibinfo{person}{Denise~R. Simmons}.}
  \bibinfo{year}{2014}\natexlab{}.
\newblock \showarticletitle{Co-curricular experiences link to nontechnical
  skill development for African-American engineers: Communication, teamwork,
  professionalism, lifelong learning, and reflective behavior skills}. In
  \bibinfo{booktitle}{\emph{2014 IEEE Frontiers in Education Conference (FIE)
  Proceedings}}. \bibinfo{pages}{1--7}.
\newblock
\urldef\tempurl%
\url{https://doi.org/10.1109/FIE.2014.7044076}
\showDOI{\tempurl}


\bibitem[Zuniga-Ruiz and Gutiérrez(2023)]%
        {chicana-math-methodology}
\bibfield{author}{\bibinfo{person}{Sandra Zuniga-Ruiz} {and}
  \bibinfo{person}{Kris~D. Gutiérrez}.} \bibinfo{year}{2023}\natexlab{}.
\newblock \showarticletitle{Pláticas as Feminista cultural practice and design
  methodology for being and becoming with mathematics}.
\newblock \bibinfo{journal}{\emph{International Journal of Qualitative Studies
  in Education}} \bibinfo{volume}{0}, \bibinfo{number}{0}
  (\bibinfo{year}{2023}), \bibinfo{pages}{1--12}.
\newblock
\urldef\tempurl%
\url{https://doi.org/10.1080/09518398.2023.2181442}
\showDOI{\tempurl}


\end{thebibliography}

\clearpage
\appendix
\section{Interview Procedure}
\subsection{Screening Survey}
\subsubsection{Email Invitation}
Dear [Member's Name], I hope this email finds you well. \avela{} is conducting research on STEM education access and the experiences of underrepresented minority students. I am reaching out to you with an invitation to participate in an interview to learn about your unique experiences not only as a student, but also as an \avela{} mentor.

If you are interested in participating, please sign up for an interview using this calendar link. 
Interviews will be conducted between 8/20/2023 and 8/27/2023.
You will be given \$20 for your participation in the interview. 

Thank you,
The \avela{} Team

\subsubsection{Informed consent}
Thank you for choosing to participate in this study, participation is completely voluntary so we appreciate your time. All information gathered from this study will remain entirely confidential. At the end of the study, you will be given a \$20 Amazon gift card for your time.

\subsubsection{Screening/Pre-interview questions}
\begin{enumerate}
    \item Did you ever qualify for free or reduced lunch in school, or are you eligible for the Pell Grant? (select yes if either is true)
        \begin{itemize}
            \item Yes
            \item No
        \end{itemize}
    \item Do you have any close family members who work, or have worked, at a technology company?
        \begin{itemize}
            \item Yes
            \item No
        \end{itemize}
    \item Did you take an alternative path to university out of secondary/high school? (running start, gap year(s), or community college)
        \begin{itemize}
            \item Yes
            \item No
        \end{itemize}
    \item Are you a first generation college student, meaning your guardian(s) did not receive a college degree?
        \begin{itemize}
            \item Yes
            \item No
        \end{itemize}
    \item In secondary/high school, did you have access to your own computer at home?
        \begin{itemize}
            \item Yes
            \item No
        \end{itemize}
    \item In secondary/high school, did you have access to after school technology programs or clubs? (robotics, computer or coding, science, etc.)
        \begin{itemize}
            \item Yes
            \item No
        \end{itemize}
    \item In secondary/high school, did you participate in an after school technology program or club? (robotics, computer or coding, science, etc.)
        \begin{itemize}
            \item Yes
            \item No
        \end{itemize}
    \item In secondary/high school, did you ever have a long-term teacher or instructor in a technology class or program who shared your same ethnic or racial identity? (Do not include substitute teachers)
        \begin{itemize}
            \item Yes
            \item No
        \end{itemize}
    \item Did any of your school teachers or staff members encourage you to participate in technology-based classes or programs?
        \begin{itemize}
            \item Yes
            \item No
        \end{itemize}
    \item Did any of your school teachers or staff members discourage you from participating in technology-based classes or programs?
        \begin{itemize}
            \item Yes
            \item No
        \end{itemize}
    \item If you have participated in a technology class or program, did the teachers or instructors speak the language you are most familiar with?
        \begin{itemize}
            \item Yes
            \item No
            \item N/A
        \end{itemize}
    \item Did anyone in your family encourage you to participate in technology-based classes or programs?
        \begin{itemize}
            \item Yes
            \item No
        \end{itemize}
    \item Did anyone in your family discourage you from participating in technology-based classes or programs?
        \begin{itemize}
            \item Yes
            \item No
        \end{itemize}
    \item In secondary/high school, did you participate in an \avela{} class or program?
        \begin{itemize}
            \item Yes
            \item No
        \end{itemize}
    \item Approximately how many \avela{} classes or programs have you taught?
        \begin{itemize}
            \item 1
            \item 2
            \item 3
            \item 4
            \item 5 or more
        \end{itemize}
    \item Which of the listed skills have you learned through \avela{}?
        \begin{itemize}
            \item Python
            \item Circuit design
            \item Data analysis and visualization 
            \item CAD Design
            \item C (or Arduino code)
            \item AI or ML
        \end{itemize}
    \item Select the AP, IB, or community college classes offered at your secondary/high school.
        \begin{itemize}
            \item Computer science
            \item Biology
            \item Environmental science
            \item Chemistry
            \item Physics
            \item Precalculus 
            \item Calculus AB
            \item Calculus BC
            \item Statistics
            \item None
        \end{itemize}
    \item Select the AP, IB, or community college classes you participated in at your secondary/high school. (select the course even if you did not take the AP exam)
        \begin{itemize}
            \item Computer science
            \item Biology
            \item Environmental science
            \item Chemistry
            \item Physics
            \item Precalculus 
            \item Calculus AB
            \item Calculus BC
            \item Statistics
            \item None
        \end{itemize}
    \item What is your age?
    \item What is your gender identity?
        \begin{itemize}
            \item Man
            \item Woman
            \item Non-binary
            \item other
        \end{itemize}
    \item What race/ethnicity best describes you?
        \begin{itemize}
            \item American Indian or Alaskan Native
            \item Black / African American
            \item Asian / Pacific Islander
            \item Hispanic / Latino
            \item White / Caucasian
            \item Other
        \end{itemize}
    \item What is the highest level of education you have completed?
        \begin{itemize}
            \item Graduated from high school/GED
            \item Associates degree
            \item Bachelors degree
            \item Masters degree
            \item PHD or MD or JD, etc.
        \end{itemize}
    \item What languages do you speak?
        \begin{itemize}
            \item English
            \item Spanish
            \item Amharic
            \item Oromo
            \item Somali
            \item Arabic
            \item Swahili
            \item Tigrinya
            \item Hausa
            \item French
            \item other
        \end{itemize}
    \item Do you identify as a person with a disability?
        \begin{itemize}
            \item Yes
            \item No
        \end{itemize}
\end{enumerate}

\section{Interview Guide}
\subsubsection{Informed Consent}
Before we begin, I’d like to remind you that your participation in this study is completely voluntary. You can choose at any point to stop the interview. Have you completed the pre-interview survey? During this interview, I want to ask you about your experience as an underrepresented minority student and technology access and education. We ask that you provide your honest opinion because it will help us understand how \avela{} has impacted you. We will also be recording the audio of this conversation to make sure we remember your ideas well. What you say will be kept confidential, not tied to your name, and only used for research. With that said, do you have any questions before we begin?

\subsubsection{Interview Questions}
1. Digital divide underlying factors
\begin{enumerate}
    \item Are you originally from the U.S. or outside the U.S.?
    \item Did you feel like you had equal access to technology as the rest of your peers? What about compared to the average Seattle student, WA state student?
        \begin{enumerate}
             \item Why not or why not?
        \end{enumerate}
    \item What factors do you believe were barriers or obstacles for you when you tried to use technologies?
        \begin{enumerate}
            \item What technologies were challenging to use?
        \end{enumerate}
    \item What factors do you believe were barriers or obstacles for you when you tried to use technologies?
        \begin{enumerate}
            \item What technologies were challenging to use?
        \end{enumerate}
    \item What factors do you believe were helpful or advantageous for you when you tried to use technologies?
        \begin{enumerate}
            \item What technologies were useful to use?
        \end{enumerate}
    \item Do you feel represented in technology spaces or careers?
        \begin{enumerate}
            \item Why or why not?
            \item If yes, can you tell me about an experience that made you feel like you were represented? (school, work, with peers, etc.)
            \item If not, can you tell me about an experience that made you feel like you were not represented? (school, work, with peers, etc.)
            \item If not, what would help make those spaces more accessible?
        \end{enumerate}
    \item Do you feel like technology is representative of your community, culture, and/or values?
        \begin{enumerate}
            \item Why or why not?
            \item Do you think you will/would design technology differently?
        \end{enumerate}
2. Community cultural wealth approach
    \item What were your motivations for obtaining higher education?
    \item What were your motivations for obtaining a degree in STEM?
    \item How did your upbringing affect your access to certain technologies?
        \begin{enumerate}
            \item Were there any parts of your religion, socio-economic status, culture, language, family ideals, etc. that affected your access to certain technologies?
        \end{enumerate}
3. How effective are \avela{} strategies
    \item Have you worked on a project with \avela{} that you believe had an impact on your community?
        \begin{enumerate}
            \item Can you describe that impact, was it an opportunity/experience that you shared with others?
        \end{enumerate}
    \item Since joining \avela{} have you received any mentorship from another \avela{} member?
        \begin{enumerate}
            \item Did you perceive this mentor as a peer, or someone near your level?
            \item What was the mentorship experience like?
            \item What did you like about this mentorship?
            \item What did you not like about this mentorship?
            \item Were you comfortable asking questions to this mentor? Why or why not?
        \end{enumerate}
    \item Since joining \avela{} have you received any technical support from another \avela{} member?
    \item Why did you join \avela{}?
        \begin{enumerate}
            \item What was appealing about this organization?
            \item What was unappealing about this organization?
            \item What were motivating factors? Family, money, community?
        \end{enumerate}
    \item Have your career plans, either long term (major, dream job, etc.) or short term (summer internship, class schedule, etc.) changed as a result of your participation in \avela{}?
    \item Through \avela{} have you been made aware of new technologies that you see yourself using either in the long term (major, dream job, etc.) or short term (summer internship, class schedule, etc.)?
        \begin{enumerate}
            \item Have you used any skills or knowledge acquired through participation in \avela{} outside of teaching an \avela{} class?
                \begin{enumerate}
                    \item Which skills or knowledge?
                    \item How did you use them?
                \end{enumerate}
        \end{enumerate}
    \item Within the context of \avela{}, can you tell me about a time when your experience with technology further ignited your interest in the tech field?
    \item Within the context of \avela{}, can you tell me about a time when your experience with technology decreased your interest in the tech field?
    \item Has participating in \avela{} ever been a burden?
        \begin{enumerate}
            \item Why or why not?
        \end{enumerate}
Closing Question
We’re down to the final question before we end the interview.
    \item Is there anything else you would like to say about your experience in \avela{} or technology access and education?
\end{enumerate}

\section{Student Feedback Surveys (Pre- and Post-class)}
\subsubsection{Pre-Survey Questions}
\begin{enumerate}
    \item What grade/year are you in? **Running Start students - For our tracking, please select your actual grade and not your year in Community College. Thanks!
        \begin{itemize}
            \item 6th
            \item 7th
            \item 8th
            \item 9th
            \item 10th
            \item 11th
            \item 12th
        \end{itemize}
    \item What school or program is hosting this class?
    \item Which class are you taking?
        \begin{itemize}
            \item Python Coding (Fractals, Machine Learning, Pong, Tetris, etc.)
            \item Fundamentals of Website Design (HTML \& CSS)
            \item Portable Door Alarm System (or other Arduino class)
            \item Analog Heart Rate Monitor
            \item Other
        \end{itemize}
    \item Before this course, how much have you learned about the topic of this class?
        \begin{itemize}
            \item This is my first time hearing about this topic
            \item I've heard of it before but don't know a lot (0-1 hour of learning)
            \item I know a little bit (2-5 hours of learning)
            \item I know a lot and still trying to learn more (6-10 hours of learning)
            \item I have a lot of experience learning and exploring this topic (10+ hours of learning)
        \end{itemize}
    \item I have taken a STEM (Science Technology Engineering or Math) class outside of school that was NOT taught by \avela{}?
        \begin{itemize}
            \item True
            \item False
        \end{itemize}
    \item From what you can remember, have you heard of any of these topics?
        \begin{itemize}
            \item Python Coding
            \item Machine Learning (ML or AI)
            \item Arduino Electronics
            \item HTML and CSS
            \item Digital Design
            \item Circuit Design
            \item Computer Aided Design (CAD)
            \item Fundamentals of Chemistry
        \end{itemize}
    \item In the future, I hope to have a career in STEM.
        \begin{itemize}
            \item Strongly disagree
            \item Disagree
            \item Neutral
            \item Agree
            \item Strongly agree
        \end{itemize}
    \item There are people I look up to or admire who work/study in STEM (Science Technology Engineering or Math). 
        \begin{itemize}
            \item Strongly disagree
            \item Disagree
            \item Neutral
            \item Agree
            \item Strongly agree
        \end{itemize}
    \item I can make a positive impact on my community through STEM (Science Technology Engineering or Math).
        \begin{itemize}
            \item Strongly disagree
            \item Disagree
            \item Neutral
            \item Agree
            \item Strongly agree
        \end{itemize}
    \item I know what types of classes to take at my school in the future to learn more about STEM (Science Technology Engineering or Math).
        \begin{itemize}
            \item Strongly disagree
            \item Disagree
            \item Neutral
            \item Agree
            \item Strongly agree
        \end{itemize}
    \item Do you have any other suggestions or comments? (how to improve the class, resources you want to see provided, hobbies or interests you have that you want us to incorporate into class, etc.)
\end{enumerate}

\subsubsection{Post-Survey Questions}
\begin{enumerate}
    \item What grade/year are you in? **Running Start students - For our tracking, please select your actual grade and not your year in Community College. Thanks!
        \begin{itemize}
            \item 6th
            \item 7th
            \item 8th
            \item 9th
            \item 10th
            \item 11th
            \item 12th
        \end{itemize}
    \item What school or program is hosting this class?
    \item Which class are you taking?
        \begin{itemize}
            \item Python Coding (Fractals, Machine Learning, Pong, Tetris, etc.)
            \item Fundamentals of Website Design (HTML \& CSS)
            \item Portable Door Alarm System (or other Arduino class)
            \item Analog Heart Rate Monitor
            \item Other
        \end{itemize}
    \item I am comfortable with the topics taught in this class.
        \begin{itemize}
            \item Strongly disagree
            \item Disagree
            \item Neutral
            \item Agree
            \item Strongly agree 
        \end{itemize}
    \item I am confident in my ability to solve engineering problems given my experience. 
        \begin{itemize}
            \item Strongly disagree
            \item Disagree
            \item Neutral
            \item Agree
            \item Strongly agree 
        \end{itemize}
    \item From what you can remember, what did you learn in this course?
        \begin{itemize}
            \item Python Coding
            \item Machine Learning (ML or AI)
            \item Arduino Electronics
            \item HTML and CSS
            \item Digital Design
            \item Circuit Design
            \item Computer Aided Design (CAD)
            \item Fundamentals of Chemistry
        \end{itemize}
    \item Course content was organized and well planned
        \begin{itemize}
            \item Strongly disagree
            \item Disagree
            \item Neutral
            \item Agree
            \item Strongly agree 
        \end{itemize}
    \item Course workload was realistic
        \begin{itemize}
            \item Strongly disagree
            \item Disagree
            \item Neutral
            \item Agree
            \item Strongly agree 
        \end{itemize}
    \item Course was engaging and encouraged participation
        \begin{itemize}
            \item Strongly disagree
            \item Disagree
            \item Neutral
            \item Agree
            \item Strongly agree 
        \end{itemize}
    \item Instructors were easy to approach and helpful
        \begin{itemize}
            \item Strongly disagree
            \item Disagree
            \item Neutral
            \item Agree
            \item Strongly agree 
        \end{itemize}
    \item Instructors were effective lecturers/demonstrators
        \begin{itemize}
            \item Strongly disagree
            \item Disagree
            \item Neutral
            \item Agree
            \item Strongly agree 
        \end{itemize}
    \item Instructors were flexible and ready to repeat things not well understood
        \begin{itemize}
            \item Strongly disagree
            \item Disagree
            \item Neutral
            \item Agree
            \item Strongly agree 
        \end{itemize}
    \item Instructors stimulated student interest
        \begin{itemize}
            \item Strongly disagree
            \item Disagree
            \item Neutral
            \item Agree
            \item Strongly agree 
        \end{itemize}
    \item Instructors effectively used time during class periods
        \begin{itemize}
            \item Strongly disagree
            \item Disagree
            \item Neutral
            \item Agree
            \item Strongly agree 
        \end{itemize}
    \item Instructors were available and helpful
        \begin{itemize}
            \item Strongly disagree
            \item Disagree
            \item Neutral
            \item Agree
            \item Strongly agree 
        \end{itemize}
    \item Instructors were patient
        \begin{itemize}
            \item Strongly disagree
            \item Disagree
            \item Neutral
            \item Agree
            \item Strongly agree 
        \end{itemize}
    \item Instructors were flexible and ready to repeat things not well understood
        \begin{itemize}
            \item Strongly disagree
            \item Disagree
            \item Neutral
            \item Agree
            \item Strongly agree 
        \end{itemize}
    \item Is the topic covered in class something you would be interested in learning more about?
        \begin{itemize}
            \item More of this!
            \item I've learned that I'm not interested in this topic
            \item In between the two
        \end{itemize}
    \item What did you like about the class?
    \item What did you dislike about the class?
    \item Do you have any comments, feedback, or critiques for either of the instructors?
    \item Do you have any other suggestions or comments? (e.g. how to improve the class, resources you want to see provided, etc.)
\end{enumerate}

\section{Interview Codebook}
The codebook is available at:  \href{https://docs.google.com/document/d/12QOSSaI59HkCZOUKrcO9BBvg0tEnwb7GPDrxN3XjrNs/edit?usp=sharing}{tinyurl.com/CHI24LBW-codebook}

\section{Participants' Demographic Information}
\begin{figure}[b]
  \centering
  \includegraphics[width=1.0\linewidth]{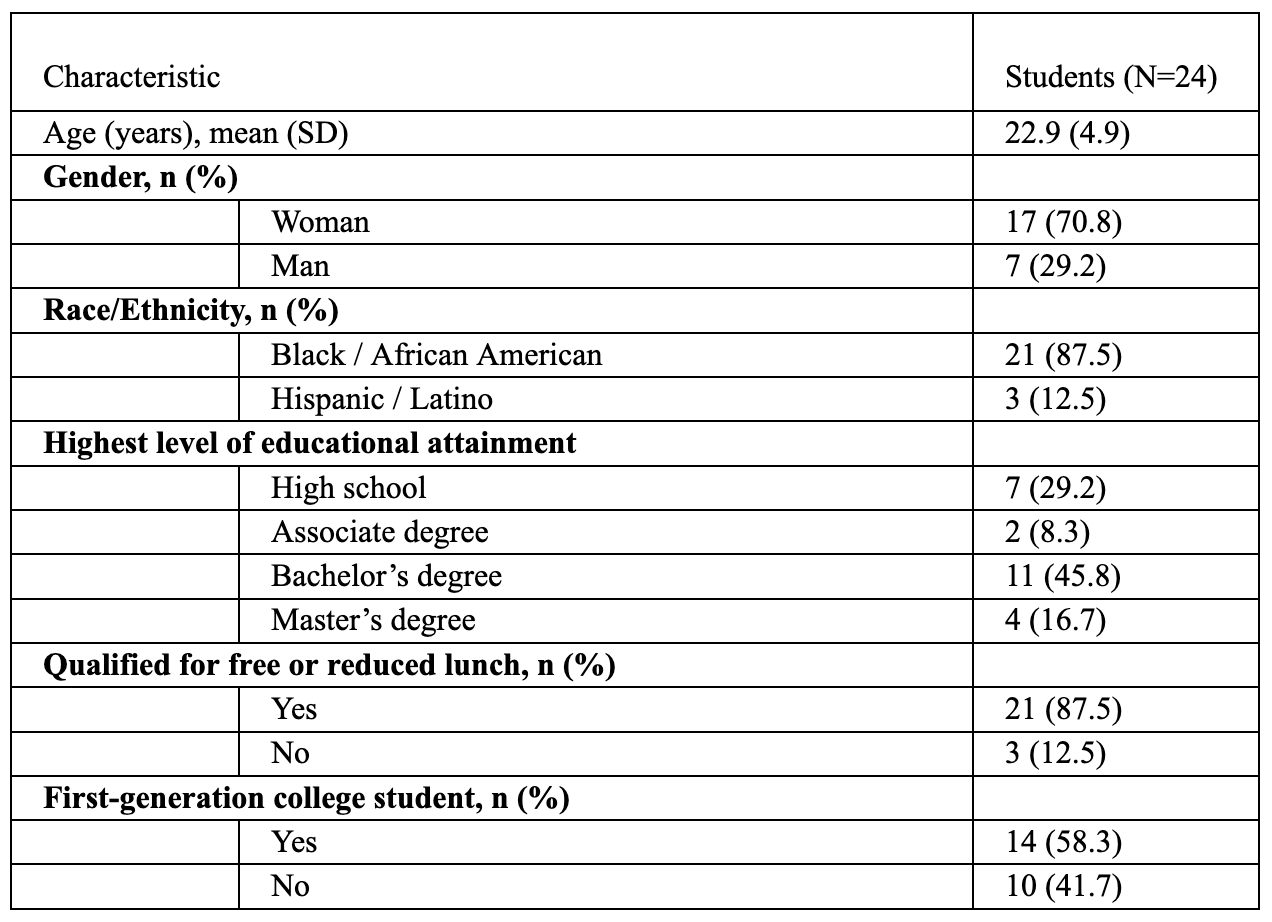}
  \caption{Demographic information of interview participants.}
  \Description{Overview of interview participants’ demographic information.}
  \label{fig:demographics}
\end{figure}

\end{document}